\documentclass[debug]{rmaa}

\usepackage{paralist}

\usepackage{psfrag,color}

\usepackage[latin1]{inputenc}

\title{TESS Data Mining and UVES Spectral Analysis of CP Stars} 

\author{
  M. E. Mac\'{i}as,\altaffilmark{1} 
  J. A. Rosales,\altaffilmark{1}{2}}

\altaffiltext{1}{Universidad Internacional de Valencia - VIU, C/Pintor Sorolla 21, 46002, Valencia, Spain.}

\altaffiltext{2}{Departamento de Astronom\'{i}a, Universidad de Concepci\'{o}n, Casilla 160-C, Concepci\'{o}n, Chile}

\shortauthor{Mac\'{i}as \& Rosales}
\shorttitle{RevMexAA Main Journal Demo Document}

\listofauthors{M. E. Mac\'{i}as \& J. A. Rosales}
\indexauthor{Mac\'{i}as, M. E.}
\indexauthor{Rosales, J. A.}

\abstract{Photometric data from TESS were analyzed to obtain the light curve, confirming that the orbital period ($P = 11.307 \pm 0.005$ days) remains constant. A wavelet analysis was applied to detect temporal anomalies. The photometric effective temperature of HD\,72968 was estimated through colorimetry, comparing it with main-sequence stars. UVES spectra were analyzed, Doppler-corrected, and normalized using Chebyshev functions. A theoretical spectral grid generated with SPECTRUM and ATLAS9 was fitted using the chi-squared method. A comparison with UVES-ESO spectra refined the temperature estimate of HD\,72968. The analysis of the spectral energy distribution (SED) enabled the derivation of stellar parameters, validating the models employed. These results provide new insights into the atmosphere and variability of HD\,72968 and motivate further spectral studies for improved stellar modeling.}

\resumen{Se analizaron datos fotom\'{e}tricos de TESS para obtener la curva de luz, confirmando que el peri\'{o}do orbital ($P = 11.307 \pm 0.005$ d\'{i}as) es constante. Se aplic\'{o} un an\'{a}lisis wavelet para detectar anomal\'{i}as temporales. La temperatura efectiva fotom\'{e}trica de HD\,72968 se estim\'{o} por colorimetr\'{i}a, compar\'{a}ndola con estrellas de la secuencia principal. Se analizaron espectros UVES corregidos por Doppler y normalizados con funciones de Chebyshev. Una cuadr\'{i}cula espectral te\'{o}rica generada con SPECTRUM y ATLAS9 fue ajustada mediante m\'{e}todo chi-cuadrado. El contraste con espectros UVES-ESO refin\'{o} la temperatura de la estrella HD\,72968. El an\'{a}lisis de la distribuci\'{o}n espectral de energ\'{i}a (SED) permiti\'{o} derivar par\'{a}metros estelares, validando los modelos empleados. Los resultados aportan nuevas perspectivas sobre la atm\'{o}sfera y variabilidad de HD\,72968, y motivan estudios espectrales m\'{a}s detallados.}

\addkeyword{TESS}
\addkeyword{UVES spectroscopy}
\addkeyword{chemically peculiar stars}
\addkeyword{spectral energy distribution}
\addkeyword{stellar parameters}

\begin{document}
% Typeset article header
\maketitle

\section{INTRODUCTION} \label{Sec:Sec. 1}

Chemically peculiar (CP) stars were first identified and classified by \citet{1897AnHar..28....1M} during spectral classification efforts in the Henry Draper Memorial project at Harvard, specifically describing the notable spectral characteristics of $\alpha^{2}$ CVn (Cor Caroli). Later, \citet{1931ApJ....73..104M} focused on Ap stars and certain manganese stars, correlating abundance anomalies with ionization temperatures and defining five groups of peculiar A stars \citep{1933PAAS....7...11M}.

Additionally, \citet{1958ApJS....3..141B} conducted a comprehensive study of 70 A-type stars with sharp or ultra-sharp lines, including subdwarfs, M giants, and S-type stars, recognizing magnetic variability as a general feature. CP stars exhibit an excess abundance of elements with $\rm{Z>26}$, sometimes exceeding two orders of magnitude compared to the solar abundance. The strong magnetic fields of these stars have been linked to these abundance anomalies, with typical horizontal field strengths reaching thousands of gauss \citep{1968BAAA...11...25J}. Diffusion processes in A-type stars \citep{1970ApJ...160..641M} suggest that CP stars make up ~10\% of early B to F main-sequence stars. \citet{1974ARA&A..12..257P} later proposed a four-group classification based on abundance anomalies:

\ \\
\begin{itemize}
\item CP1: Am/Fm stars, showing abnormal Ca II and/or Sc II surface anomalies, overabundance of iron-peak and heavier elements, no strong global magnetic fields, spectral types F0 to A0, temperatures 7000 $\leq T \leq$ 10000\,K.
\item CP2: Ap stars, enhanced Si, Cr, Sr, Eu, or rare-earth elements, with strong magnetic fields, spectral types F4 to B6, temperatures 8000 $\leq T \leq$ 15000\,K.
\item CP3: HgMn stars, enriched Hg II ($\mathrm{\lambda}$\,3984), Mn, and other heavy elements, mostly non\textrm{-}magnetic, spectral types A0 to B6, temperatures 10000 $\leq T \leq$ 15000\,K.
\item CP4: He I-weak stars, weak helium lines, some with detectable magnetic fields, spectral types B8 to B2, temperatures 13000 $\leq T \leq$ 20000\,K.
\end{itemize}

\ \\
\noindent
A later refinement added two additional subgroups: $\lambda$ Bootis stars, with weak Mg II and metal-deficient spectra (F0 to A0, temperatures 7500 $\leq T \leq$ 9000\,K), and He-rich stars, showing enhanced He I, partially strong magnetic fields (B2, temperatures $\mathrm{20000 \leq T \leq 25000 ~K}$) \citep{1958SvA.....2..151P, 1996Ap&SS.237...77S,2004IAUS..224..443P}. Thus, six CP subclasses exist today, expanding the original four groups.

CP stars display ultraviolet flux deficiencies compared to normal stars with similar Balmer jumps \citep{1998A&AS..129..463C}. The hottest CP stars exhibit unusually strong Mn, Si, and Hg lines, whereas cooler ones show enhanced Si, Cr, Sr, and Eu. Spectral type Ap (Mn, Hg) stars have temperatures of 10000 \textrm{-} 15000 K, rotation speeds of 30 $\mathrm{km\,s^{-1}}$, weak magnetic fields, and normal binary occurrence rates; Ap (Sr, Eu) stars have temperatures of 8000 \textrm{-} 12000 K, rotation speeds of 30 $\mathrm{km\,s^{-1}}$, magnetic fields of $10^3$\textrm{-}$10^4$ gauss, and lower binary occurrence rates \citep{2000asqu.book.....C}.

Ap stars extend into early B and F types, covering spectral ranges B8\textrm{-}F5 in a region of the Hertzsprung-Russell diagram relatively free of astrophysical complications like strong convection or significant mass loss. However, 10\% of these stars exhibit strong line anomalies in their spectra, forming the CP group \citep{2011BAAA...54..167J}.

According to VSX-AAVSO\footnote{\url{https://www.aavso.org/vsx/}}, HD\,72968 is classified as an Alpha2 Canum Venaticorum variable (ACV), designated ID 000-BBP-746, with equatorial coordinates $\alpha_{2000}=08:35:28.2$ and $\delta_{2000}=-07:58:56.2$. Its magnitude varies between $5.72 \leq \mathrm{V\,(mag)} \leq 5.74$, with a period of $\mathrm{P=11.305\,d}$. Using differential Str\"{o}mgren $uvby$ observations, \citet{2005A&A...435.1099A} classified it as a magnetic Chemically Peculiar (mCP) star. Its distance is d = 107.531 pc, based on Gaia DR3\footnote{\url{https://gea.esac.esa.int/archive/}}.

\section{Photometric analysis}\label{Sec:Sec. 2}

With the aim of retrieving and analyzing scientific data collected by space telescopes, such as the \textquotedblleft{Transiting Exoplanet Survey Satellite\textquotedblright}, commonly known as TESS\footnote{\url{https://tess.mit.edu/}}, an essential resource for the scientific community interested in astronomical data from these missions is the \textquotedblleft{Mikulski Archive for Space Telescopes\textquotedblright} (MAST)\footnote{\url{https://mast.stsci.edu/portal/Mashup/Clients/Mast/Portal.html}}. By relating the photometric study of TESS data with UVES spectroscopy, we aim to resolve key aspects for the accurate estimation of values and error margins of the physical parameters of HD\,72968. \citet{2024NewA..10702135A} states that by combining spectroscopic data with photometric information in the light curve analysis, future observations may provide even more precise findings. The simultaneous analysis of spectroscopic and photometric data of the system has the potential to yield significantly refined results.

%##################################################################### %##################################################################### \section{Photometric Analysis}

\begin{figure}[] 
\centering 
\includegraphics[trim=2.0cm 0.0cm 2.0cm 2.0cm, width=0.6\textwidth,angle=0]{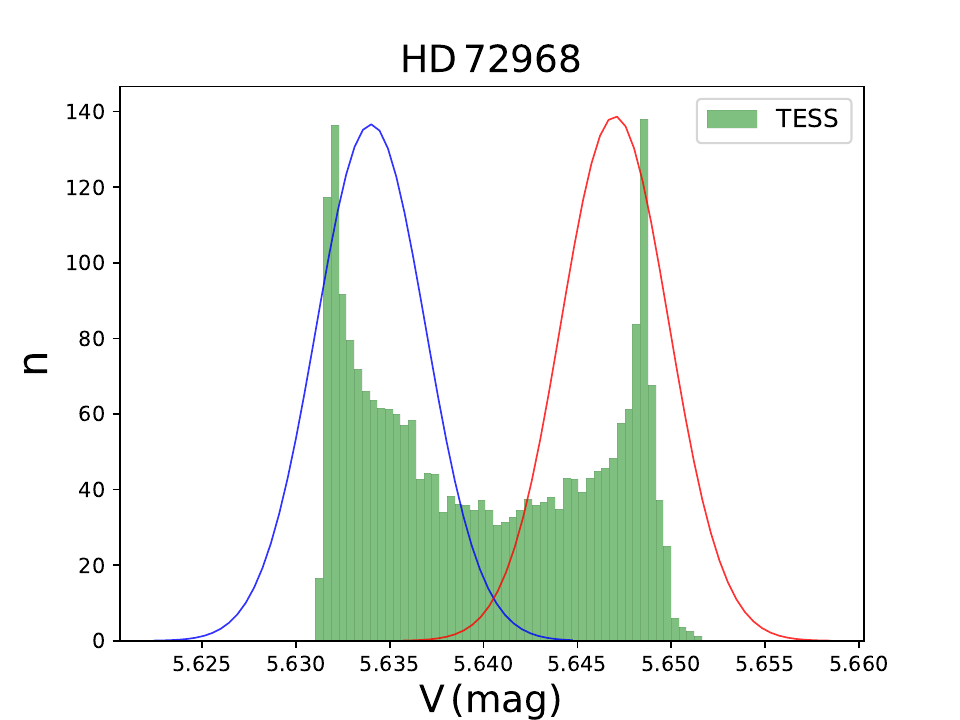} 
\vskip 0.25cm 
\caption{Magnitude histogram for HD\,72968 using 50 bins for 10,703 TESS data points, showing a bimodal relationship possibly caused by a pulsation affecting its brightness.} 
\label{fig:Fig. 2.1} 
\end{figure}

A dataset of 10703 high-precision photometric measurements in the $V$ band, with a cadence of approximately 120 seconds over a 25-day sampling segment, was obtained using the Transiting Exoplanet Survey Satellite (TESS). These data are recorded in the TESS Barycentric Julian Date (BJD) and were subsequently corrected to Heliocentric Julian Date using the following equation:

\begin{equation} \mathrm{BJD=HJD-2457000.0},
\label{eq: eq. 2.1} \end{equation} \vskip 0.25cm

\begin{equation} \mathrm{V_\mathrm{{TESS}} = -2.5 \cdot log_{10}(\mathrm{F_{TESS}}) + Z_{p}}, \label{eq: eq. 2.2} \end{equation} \vskip 0.25cm

\noindent
where $\mathrm{{V_{TESS}}}$ corresponds to the corrected apparent magnitude from TESS, $\rm{F_{TESS}}$ represents the observed flux in ($\mathrm{e^{-}s^{-1}}$), and $Z_{p}$ corresponds to the correction factor or TESS zero point of $\mathrm{Z_{p}=20.44}$ (mag). Additionally, for this purpose, a histogram analysis has been performed in which we grouped the data into 50 bins. We note that the maximum apparent magnitude of the star HD\,72968 is $V= 5.6310 \pm 0.0002\,(\mathrm{mag})$, with a mean of $\bar{x}= 5.6399 \pm  0.0002\,(\mathrm{mag})$ (see Fig. \ref{fig:Fig. 2.1}).

In order to better understand the observed photometric variation in this star, we fitted two Gaussian distributions to attempt to explain how the brightness of this star varies. The first fit follows a Gaussian function of the form $\mathrm{f(x|\mu,\sigma)=1/\sigma\sqrt{2\pi}e^{{-1/2}({x-\mu}/\sigma)^{2}} }$, where x is the random variable representing the magnitudes, $\mathrm{\mu}$ is the mean of the distribution, and $\mathrm{\sigma}$ is the standard deviation of the distribution, resulting in a mean value of $\mathrm{\mu = 5.634}$ (mag) and a standard deviation of $\mathrm{\sigma = 0.00292}$. 

The second distribution is of the same type, yielding a mean of $\mathrm{\mu = 5.647}$ (mag) and a standard deviation of $\mathrm{\sigma= 0.002875}$. Therefore, this shows that the distribution is bimodal but not normal, covering 95\% of the photometric data. Additionally, the minimum apparent magnitude is $\mathrm{m_{min}}= 5.6516 \pm 0.0002\,\mathrm{(mag)}$ (see Fig. \ref{fig:Fig. 2.1}).

\begin{figure}[!h] 
\centering 
\includegraphics[trim=0.2cm 0.1cm 0.3cm 0.2cm, width=0.45\textwidth,angle=0]{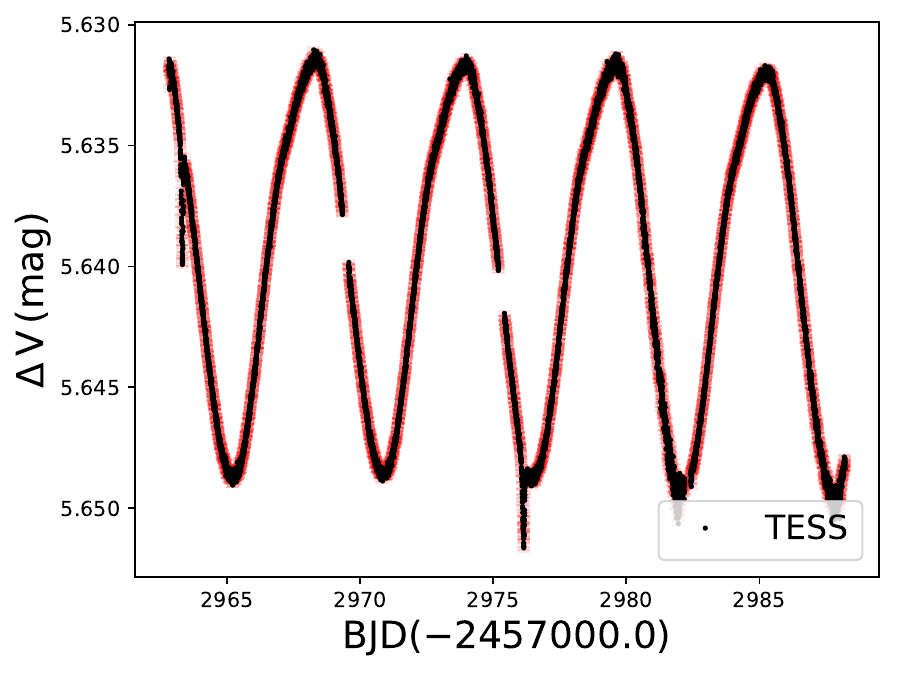} 
\includegraphics[trim=0.2cm 0.1cm 0.3cm 0.2cm, width=0.45\textwidth,angle=0]{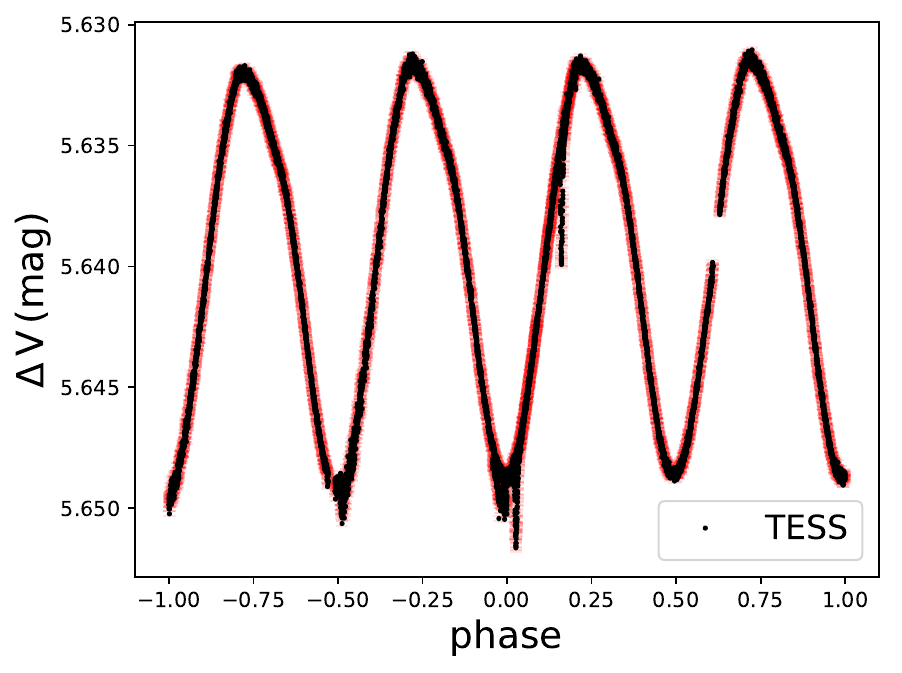} 
\vskip 0.25cm 
\caption{(Left) TESS photometric flux transformed to magnitude according to Eq. 
\ref{eq: eq. 2.2}, distributed over BJD(-2457000.0). (Right) Light curve for HD\,72968, according to Eq. 
\ref{eq: eq. 2.3}.} 
\label{fig:Fig. 2.2} 
\end{figure}

Subsequently, we conducted a photometric analysis of the TESS light curve, considering all 10703 quality TESS data points. The dataset was analyzed using the Phase Dispersion Minimization (\texttt{PDM-IRAF}) algorithm of \citep{1978ApJ...224..953S}\footnote{IRAF is distributed by the National Optical Astronomy Observatories, operated by the Association of Universities for Research in Astronomy, Inc., under cooperative agreement with the National Science Foundation.}. The minimum and maximum values were determined through light curve analysis to establish the correct period. In this study, we examined data covering periods ranging from 1 to 20 days. Once the period was identified within the range of 10.9 to 11.7 days, decimal adjustments were made in increments of $\pm 0.1$ days until achieving optimal precision in the period value. Consequently, analyzing the light curve of HD\,72968, we obtained an orbital period of $\mathrm{P = 11.307 \pm 0.005,\mathrm{d}}$, thus corroborating the previously reported value of 11.305 days for the orbital period of the same chemically peculiar star, according to the photometric light curve studies by \citet{1978A&A....62..199M} and \citet{1998A&AS..128..245A}. Therefore, we determined the following ephemeris for the light curve, which will be used for further analysis in the document:

\begin{equation} \mathrm{HJD_{min} = 2459976.4413447+ 11.307(5) \cdot E}, \label{eq: eq. 2.3} \end{equation} \vskip 0.25cm

\noindent
where $\mathrm{HJD_{min}}$ corresponds to the Heliocentric Julian Date, the time HJD = 2459976.4413447 corresponds to the reference epoch, P is the orbital period of the star HD\,72968, and E is the number of complete epochs elapsed since 2459976.4413447, which is the reference epoch.

For the graphical representation shown in Figure \ref{fig:Fig. 2.2}, the advanced software PHOEBE\footnote{\url{http://phoebe-project.org/}} was used. This software utilizes previously obtained parameters, such as the period $\mathrm{P} = 11.307 \pm 0.005,\mathrm{d}$ and the time in BJD 2976.14736220 of the minimum apparent magnitude. This allows for the generation of the magnitude-phase plot, centered at 0, with positive and negative extremes of 1 and -1, respectively.

%##################################################################### %##################################################################### %##################################################################### \clearpage 
\section{Wavelet Analysis of Data}

For a more localized and adaptive representation of the characteristics of the photometric data, we decided to use the mathematical technique that employs wavelet functions to analyze and process signals and data. In brief, unlike other frequency analysis techniques, such as the Fourier transform, which uses sinusoidal functions, wavelet data analysis provides a more localized and adaptive representation of a signal's characteristics. Wavelet functions are small waves (wavelets) that are localized in both time and frequency.

\begin{figure}[]
\centering 
\includegraphics[trim=0.2cm 0.1cm 0.3cm 0.2cm, width=1\textwidth,angle=0]{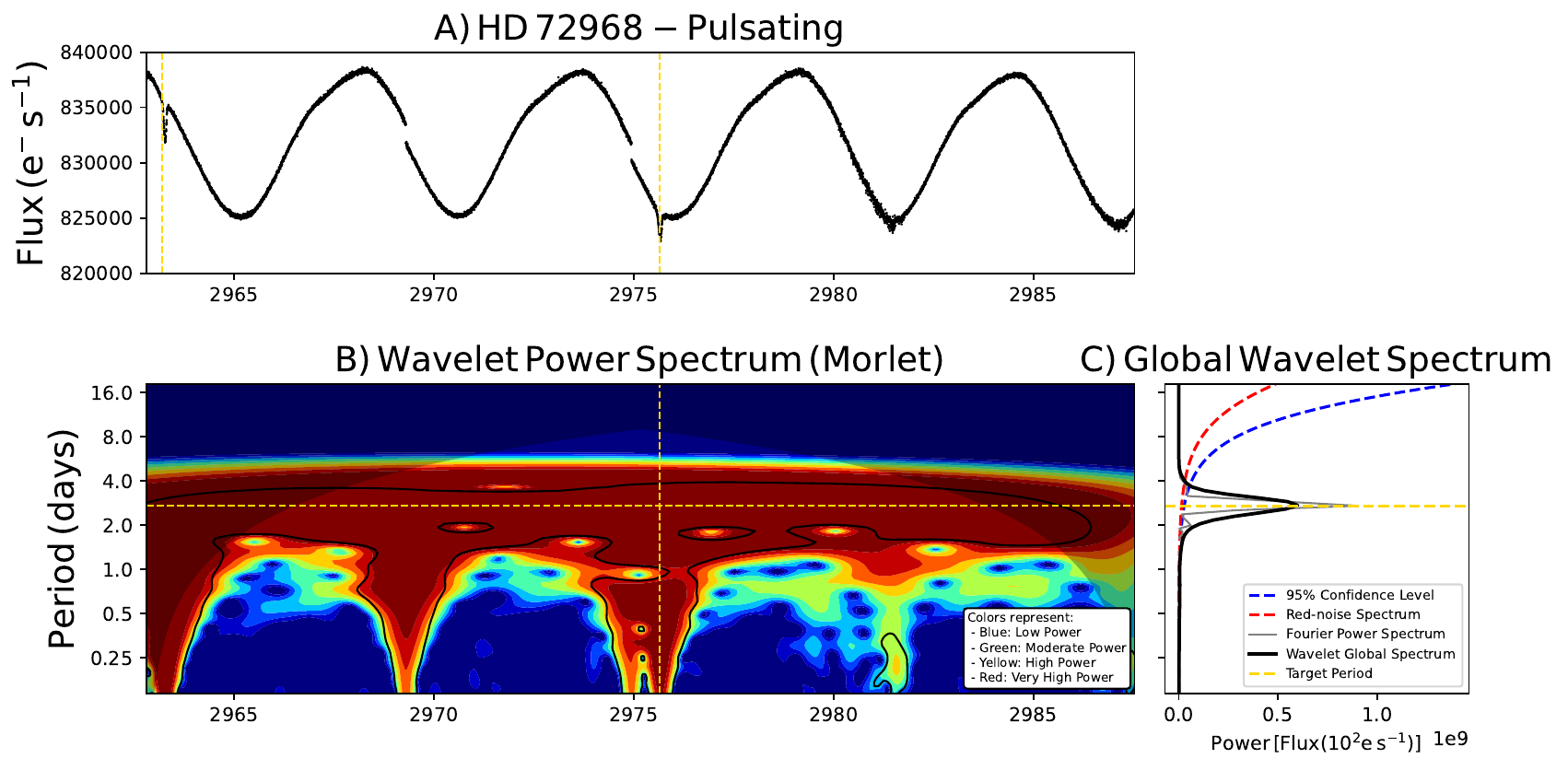} 
\vskip 0.25cm 
\caption{Plot A: Temporal evolution of the observed flux (flux vs. time), with the detected variabilities marked by dashed yellow lines. Plot B: Normalized wavelet spectrum. The gray area represents the cone of influence. The horizontal dashed yellow line corresponds to the detected period $P_2$ of the star, $P_2 = 2.7$ days. The vertical dashed yellow line near $\mathrm{BJD = 2976.15}$ highlights an anomaly in HD\,72968 as it enters its minimum magnitude. The black contour lines enclosing various regions of the plot indicate a 95\% statistical significance level. A jet color palette is used, where blue represents low wavelet power, red indicates high wavelet power, and intermediate power values are shown in green and yellow, respectively. Plot C: Global wavelet power spectrum. The dashed black line shows the average power spectrum, with a prominent peak at $P_2 = 2.7$ days. The solid gray line represents the Fourier Transform, included for comparison with the wavelet analysis. Dashed red lines indicate the noise level, with the average power peak lying above it, suggesting the presence of a real signal. The dashed blue lines represent the 95\% significance threshold; similarly, the average power peak lies above this threshold, providing additional evidence for a real oscillation. Dashed yellow lines mark the period of interest.} 
\label{fig:Fig. 2.3} 
\end{figure}

They are generated from scaling functions that satisfy the recursion relation \citet{2004MWRv..132.1220A}, presented in equation \ref{eq: eq. 2.4}. In brief, this transform operates by using functions called wavelets, which are small waves that shift and scale to adapt to different parts of the signal, in this case, to frequencies corresponding to the star's period.

\begin{equation} \mathrm{\phi(x)=\sum\limits_{k}^{}c_k\phi (2x-k)}, \label{eq: eq. 2.4} \end{equation} \vskip 0.25cm

\noindent
where $c_k$ is a finite set of filter coefficients. For the analysis of 10,703 TESS data points, the normalized wavelet power spectrum was used, employing the Morlet wavelet ($\omega = 6$), which is a complex exponential (Fourier) wave given by equation \ref{eq: eq. 2.5}.

\begin{equation} \mathrm{\psi(t) = \exp\left(i\omega_{0}t\right) \exp\left(-\frac{t^2}{2\sigma^2}\right)}, \label{eq: eq. 2.5} \end{equation} \vskip 0.25cm

where $\mathrm{\omega_{0}}$ is the frequency and $\sigma$ is a support measure for the wavelet. For the Morlet mother wavelet, the base scale parameters were adjusted to $s_0 = 60 * dt$, the scale factor $dj = 1 / 12$, the total number of scales $J = 7 / dj$, and $\alpha = 0.77$. The normalized Morlet wavelet power spectrum is presented in Figure \ref{fig:Fig. 2.3}.

Thus, for a total of 10703 photometric data points provided by TESS, with a cadence of 3.33 minutes and a $\mathrm{T_{0}=2962.8045771}$, we performed wavelet data analysis by applying the wavelet transform. This involves shifting and scaling the mother wavelet function to adapt to different parts of the signal, as mentioned earlier. This shifting or translation moves the wavelet function along the time axis, allowing different parts of the signal to be analyzed at different moments. Meanwhile, the scaling allows the wavelet function's scale to be adjusted to analyze the signal at different frequencies. Due to the pulsating nature of this star, as it belongs to the CP family, it was necessary to consider a function with a higher frequency capable of capturing fine details in the signal. It is worth noting that this could have caused the loss of information about low-frequency components. Our mother wavelet function is of order 6, and we have considered 12 sub-octaves per octave. This means that one frequency interval is double the frequency of another interval. Therefore, when referring to "octaves" in the context of wavelets, we are discussing scales that vary in frequency logarithmically. Each octave represents a frequency range that is double that of the previous octave. The term "sub-octave" refers to frequency intervals that are smaller than an octave. Instead of doubling the frequency, a sub-octave could represent a fraction of the frequency of the previous octave. This allows for a more detailed analysis of low frequencies without covering as broad a range as a full octave.

Once a 30th-degree polynomial was fitted within the wavelet function, we noticed the existence of a new period within the light curve that had not been previously detected using the PDM-\texttt{IRAF} task. This period is $\mathrm{P_{2}= 2.7,\mathrm{d}}$ and is possibly related to internal pulsations of the star (see Fig. \ref{fig:Fig. 2.3}-B,C). Additionally, we observed two anomalies or changes in the morphology of the light curve at $\mathrm{BJD= 2976.15}$ and $\mathrm{BJD= 2963.31}$, where the star's brightness drops over a short period. We speculate that during pulsation, the star ejects a small amount of material, causing a brief eclipse at those reported moments. Furthermore, starting from $\mathrm{BJD= 2980}$, we noticed that during the star's minima, a variation in the morphology appeared, which is more extended and seems to be remnants of the second mini-eclipse or anomaly corresponding to $\mathrm{BJD= 2963.31}$ (see Fig. \ref{fig:Fig. 2.3}-A).

\newpage
\subsection{Colorimetry Method}\label{Sec:Sec. 2.1}

\begin{figure}[]
\centering
\includegraphics[trim=0.0cm 0.3cm 0.0cm 0.2cm,clip,width=0.75\textwidth,angle=0]{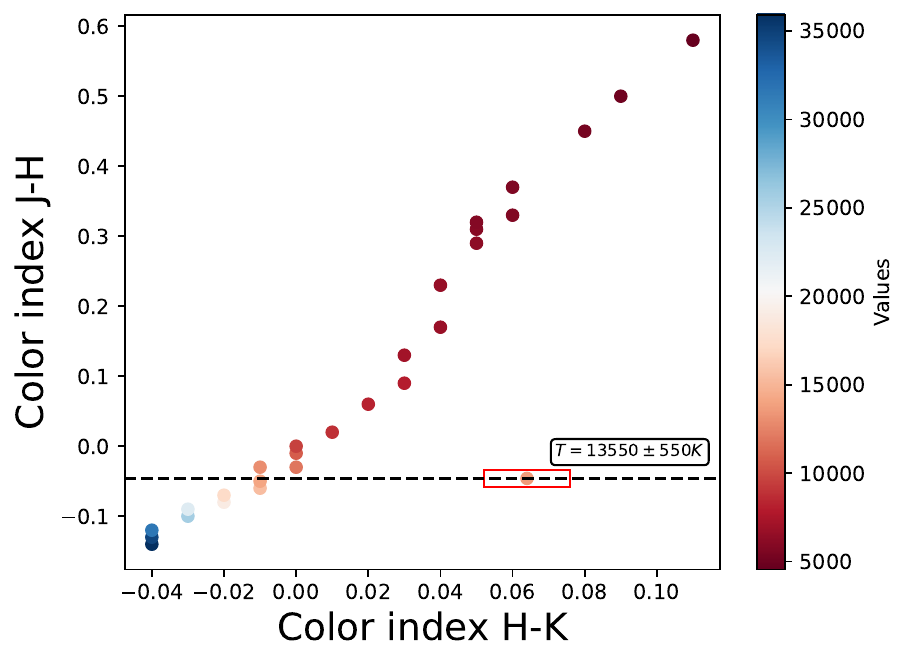}
\vskip 0.25cm
\caption{Color-temperature comparison of HD\,72968 with respect to 27 spectral types of main sequence stars.}
\label{fig:Fig. 2.4}
\end{figure}

We conducted a search for photometric data provided by ExoFOP-TESS \citep{2019AAS...23314009A} for other bands and summarized some of these in Table \ref{Tab: Tab. 2.1}. In the color-color graphical representation shown in Figure \ref{fig:Fig. 2.4}, we have plotted 27 spectral type variants spanning a temperature range from 35900 K to 4560 K \citep{2000asqu.book.....C}. The color difference was $\mathrm{J-H}=-0.046 \pm 0.069$ (mag) and $\mathrm{H-K}=0.064 \pm 0.058$ (mag), giving us clues that the star fluctuates within a fairly wide temperature range (see Table \ref{Tab: Tab. 2.2}). Placing it for the $\mathrm{J-H}$ difference at $\mathrm{13550 \pm 550,\rm{K}}$, while for the $\mathrm{H-K}$ difference, it is $\mathrm{5430 \pm 380,\rm{K}}$.

From the photometry conducted for the star HD\,72968, an initial estimate of its effective temperature was obtained, which was 10,700 K according to UBV data \citep{1974A&A....37..367H}. By plotting the color-color graph shown in Figure \ref{fig:Fig. 2.4} with the photometric data $\mathrm{J - H ,and, H - K}$ according to their position relative to main sequence stars, an estimated effective temperature of $\mathrm{13550 \pm 550 ,\mathrm{K}}$ was obtained. It is important to note that the difference between main sequence stars and chemically peculiar stars is that in chemically enriched stars, according to \citet{2005A&A...435.1099A}, anomalous photospheric abundances are produced by hydrodynamic processes, particularly radiative diffusion and gravitational settling in radiative envelopes containing strong magnetic fields.

This significant discrepancy between the initial estimate based on UBV data and the result obtained using data from \citet{2000asqu.book.....C} is due to various factors that affect information acquisition, such as different types of filters and bands, instrumental calibrations, instrument sensitivity due to constant technological updates, atmospheric correction, and reddening correction. This highlights the importance of considering different data sources and methods when analyzing the effective temperature of the star.

\begin{table}[!ht]
\caption{Photometric summary for the mCP star HD\,72968 in different bandwidths.}
\normalsize
\begin{center}
\resizebox{0.45\textwidth}{2.9cm}{
\begin{tabular}{lccrc}
\hline
\noalign{\smallskip}
\textrm{Band} & \textrm{Value (mag)} & \textrm{error (mag)} & \textrm{$\lambda_\textrm{eff}$ (\AA{})}\\
\hline
\hline
\textrm{TESS}           & \textrm{5.744} &\textrm{0.007} & \textrm{7452.64}     \\
\textrm{ExoFOP-B}       & \textrm{5.697} &\textrm{0.022} & \textrm{4400}        \\
\textrm{ExoFOP-V}       & \textrm{5.733} &\textrm{0.023} & \textrm{5500}        \\
\textrm{Gaia}           & \textrm{5.714} &\textrm{0.001} & \textrm{5857.56}     \\
\textrm{2MASS-J}        & \textrm{5.708} &\textrm{0.029} & \textrm{12350}       \\
\textrm{2MASS-H}        & \textrm{5.754} &\textrm{0.040} & \textrm{16620}       \\
\textrm{2MASS-K}        & \textrm{5.690} &\textrm{0.018} & \textrm{21590}       \\
\textrm{WISE1}          & \textrm{5.720} &\textrm{0.146} & \textrm{33526}       \\
\textrm{WISE2}          & \textrm{5.634} &\textrm{0.053} & \textrm{46028}       \\
\textrm{WISE3}          & \textrm{5.808} &\textrm{0.016} & \textrm{115608}      \\
\textrm{WISE4}          & \textrm{5.760} &\textrm{0.047} & \textrm{220883.83}   \\
\hline
\end{tabular}
}
\end{center}
\label{Tab: Tab. 2.1}
\end{table}

\begin{table}[]
\normalsize
\caption{ Intrinsic Colors and Effective Temperatures for the Main Sequence (Class V)$^{a}$.}
\begin{center}
\begin{tabular}{lcccccccr}
\hline
\noalign{\smallskip}
\textrm{Spectral type}  & \textrm{ V-K}  & \textrm{J-H}   &  \textrm{H-K}  &  \textrm{K-L}  &  \textrm{K-L'} & \textrm{K-M}  & \textrm{$T_{efl}$}\\
\hline
\hline
\textrm{O9}             & \textrm{-0.87} & \textrm{-0.14} & \textrm{-0.04} & \textrm{-0.06} & \textrm{}      & \textrm{}     & \textrm{35900}    \\
\textrm{09.5}           & \textrm{-0.85} & \textrm{-0.13} & \textrm{-0.04} & \textrm{-0.06} & \textrm{}      & \textrm{}     & \textrm{34600}    \\
\textrm{BO}             & \textrm{-0.83} & \textrm{-0.12} & \textrm{-0.04} & \textrm{-0.06} & \textrm{}      & \textrm{}     & \textrm{31500}    \\
\textrm{B1}             & \textrm{-0.74} & \textrm{-0.10} & \textrm{-0.03} & \textrm{-0.05} & \textrm{}      & \textrm{}     & \textrm{25600}    \\
\textrm{B2}             & \textrm{-0.66} & \textrm{-0.09} & \textrm{-0.03} & \textrm{-0.05} & \textrm{}      & \textrm{}     & \textrm{22300}    \\
\textrm{B3}             & \textrm{-0.56} & \textrm{-0.08} & \textrm{-0.02} & \textrm{-0.05} & \textrm{}      & \textrm{}     & \textrm{19000}    \\
\textrm{B4}             & \textrm{-0.49} & \textrm{-0.07} & \textrm{-0.02} & \textrm{-0.05} & \textrm{}      & \textrm{}     & \textrm{17200}    \\
\textrm{B5}             & \textrm{-0.42} & \textrm{-0.06} & \textrm{-0.01} & \textrm{-0.04} & \textrm{}      & \textrm{}     & \textrm{15400}    \\
\textrm{B6}             & \textrm{-0.36} & \textrm{-0.05} & \textrm{-0.01} & \textrm{-0.04} & \textrm{}      & \textrm{}     & \textrm{14100}    \\
\textrm{B7}             & \textrm{-0.29} & \textrm{-0.03} & \textrm{-0.01} & \textrm{-0.04} & \textrm{}      & \textrm{}     & \textrm{13000}    \\
\textrm{B8}             & \textrm{-0.24} & \textrm{-0.03} & \textrm{0.00}  & \textrm{-0.04} & \textrm{}      & \textrm{}     & \textrm{11800}    \\
\textrm{B9}             & \textrm{-0.13} & \textrm{-0.01} & \textrm{0.00}  & \textrm{-0.03} & \textrm{}      & \textrm{}     & \textrm{10700}    \\
\textrm{AO}             & \textrm{0.00}  & \textrm{0.00}  & \textrm{0.00}  & \textrm{0.00}  & \textrm{0.00}  & \textrm{0.00} & \textrm{9480}     \\
\textrm{A2}             & \textrm{0.14}  & \textrm{0.02}  & \textrm{0.01}  & \textrm{0.01}  & \textrm{0.01}  & \textrm{0.01} & \textrm{8810}     \\
\textrm{AS}             & \textrm{0.38}  & \textrm{0.06}  & \textrm{0.02}  & \textrm{0.02}  & \textrm{0.02}  & \textrm{0.03} & \textrm{8160}     \\
\textrm{A7}             & \textrm{0.50}  & \textrm{0.09}  & \textrm{0.03}  & \textrm{0.03}  & \textrm{0.03}  & \textrm{0.03} & \textrm{7930}     \\
\textrm{FO}             & \textrm{0.70}  & \textrm{0.13}  & \textrm{0.03}  & \textrm{0.03}  & \textrm{0.03}  & \textrm{0.03} & \textrm{7020}     \\
\textrm{F2}             & \textrm{0.82}  & \textrm{0.17}  & \textrm{0.04}  & \textrm{0.03}  & \textrm{0.03}  & \textrm{0.03} & \textrm{6750}     \\
\textrm{F5}             & \textrm{1.10}  & \textrm{0.23}  & \textrm{0.04}  & \textrm{0.04}  & \textrm{0.04}  & \textrm{0.02} & \textrm{6530}     \\
\textrm{F7}             & \textrm{1.32}  & \textrm{0.29}  & \textrm{0.05}  & \textrm{0.04}  & \textrm{0.04}  & \textrm{0.02} & \textrm{6240}     \\
\textrm{GO}             & \textrm{1.41}  & \textrm{0.31}  & \textrm{0.05}  & \textrm{0.05}  & \textrm{0.05}  & \textrm{0.01} & \textrm{5930}     \\
\textrm{02}             & \textrm{1.46}  & \textrm{0.32}  & \textrm{0.05}  & \textrm{0.05}  & \textrm{0.05}  & \textrm{0.01} & \textrm{5830}     \\
\textrm{04}             & \textrm{1.53}  & \textrm{0.33}  & \textrm{0.06}  & \textrm{0.05}  & \textrm{0.05}  & \textrm{0.01} & \textrm{5740}     \\
\textrm{06}             & \textrm{1.64}  & \textrm{0.37}  & \textrm{0.06}  & \textrm{0.05}  & \textrm{0.05}  & \textrm{0.00} & \textrm{5620}     \\
\textrm{KO}             & \textrm{1.96}  & \textrm{0.45}  & \textrm{0.08}  & \textrm{0.06}  & \textrm{0.06}  & \textrm{-0.01}& \textrm{5240}     \\
\textrm{K2}             & \textrm{2.22}  & \textrm{0.50}  & \textrm{0.09}  & \textrm{0.07}  & \textrm{0.07}  & \textrm{-0.02}& \textrm{5010}     \\
\textrm{K4}             & \textrm{2.63}  & \textrm{0.58}  & \textrm{0.11}  & \textrm{0.09}  & \textrm{0.10}  & \textrm{-0.04}& \textrm{4560}     \\
\hline
\end{tabular}
\end{center}
NOTE: $^{a}$Colors provided in the Johnson-Glass system, as established by \citep{1988PASP..100.1134B}. The references used are: O, B, [2]; A, F, O, K, [1]; K, M, [3]. K-M from [2] was not used due to a large discrepancy compared to \citep{1988PASP..100.1134B}. Approximate uncertainties (one standard deviation): $\pm$0.02 (O-K); $\pm$0.03 (M).  
$^{b} T_{{eff}}$ is an average of values from the following sources: for O, B, [4]; for B, A, F, O, K, [5]; for B, O, K, [6]; for A, F, [7]; for A, F, O, K, [8]; for A, F, O, [9]; for O, K, [10]; for K, M, [3]; for M, [11], [7], [12]. Approximate uncertainties (one standard deviation): $\pm$1000 K (O9-B2); $\pm$250 K (B3-B9); $\pm$100 K (A0-M6).
\label{Tab: Tab. 2.2}
\end{table}

\section{Spectroscopy}\label{Sec:Sec. 3}

The corrected and wavelength-calibrated spectrum was obtained from the European Southern Observatory (ESO) database\footnote{\url{https://www.eso.org/public/}} for the star HD\,72968. The processing of the obtained spectra was carried out using \texttt{IRAF}; the spectra were trimmed and continuum-normalized using the \textquotedblleft{chebyshev\textquotedblleft} fitting function (Chebyshev polynomial) and were not flux-calibrated. \citet{2023A&A...670A..94R} states, \textquotedblleft{Flux calibration of the spectra is not necessary as it does not affect line strength measurements and radial velocities\textquotedblright}. For the chemical composition analysis of the chemically peculiar star HD\,72968, the spectrum presented in Table \ref{Tab: Tab. 2.3} was used.

\vskip 0.25cm
\begin{table}[h]
\caption{Summary of the observed spectrum. N is the number of extracted spectra, HJD is the reference system used to specify the observation time, and $\phi_0$ refers to the orbital period determined using equation 3. S/N stands for signal-to-noise ratio.}
    \label{Tab: Tab. 2.3}
    \centering
    \begin{tabular}{c c c c c c c}
         \hline
         Date & Instrument & N & Exposure time (s) & HJD & $\mathrm{\phi_0}$ & S/N\\
         \hline
         \hline
         20-12-2005 & UVES & 1 & 37.5014 & 2403724.74899141 & 0.948468 & 150.7\\
         \hline
    \end{tabular}
\end{table}

After normalization, the Doppler effect correction was applied by modifying the radial velocity ($\mathrm{km\,s^{-1}}$) of the extracted spectra in \texttt{IRAF}.  

%Regarding the Doppler effect in stars, \citet{1974TESIS} stated, \textquotedblleft Stars move in all directions throughout the universe, and the Doppler effect can be observed. Sometimes, a blueshift is present, while in other cases, a redshift occurs. Stars that are barely visible tend to shift towards the red\textquotedblright.  

In addition to the Doppler effect correction, chemical element identification was performed in \texttt{IRAF} for absorption lines by specifying the wavelength in angstroms. Using the VizieR astronomical catalog\footnote{\url{http://vizier.cds.unistra.fr/}}, the main chemical elements associated with the A2p spectral type of HD\,72968 were identified for use in the present investigation.

\vskip 0.25cm
\begin{table}[h]
\caption{Chemical elements identified in the star HD\,72968 according to information obtained from VizieR, where the determined chemical element, wavelength in angstroms, spectral type range, and equivalent width (EW) of the absorption lines with the associated error ($\pm$ Rms) are represented.}
\centering
\begin{tabular}{l c l c}
\hline
\textrm{Chemical element} & \textrm{Wavelength (\AA{})} & \textrm{Spectral type} & \textrm{EW $\pm$ Rms}\\
\hline
\hline
\textrm{Sr\,II}     &  \textrm{4077.7140} &  \textrm{A0V-M2Ia} & 0.914 $\pm$ 0.098 \\
\textrm{H\,$\delta$}&  \textrm{4101.7370} &  \textrm{O9V-M2Ia} & 2.506 $\pm$ 0.019\\
\textrm{Cr\,II}     &  \textrm{4824.1300} &  \textrm{A0-A3V;A0-G0IV;A0-G0Ia} & $0.259 \pm 0.033$ \\
\textrm{H\,$\beta$}      &  \textrm{4861.3320} &  \textrm{O9V-M2Ia} & $11.720 \pm 0.060$ \\
\hline
\end{tabular}
\label{Tab: Tab. 2.4}
\end{table}

\begin{figure}[!h]
\centering
\includegraphics[width=0.8\textwidth]{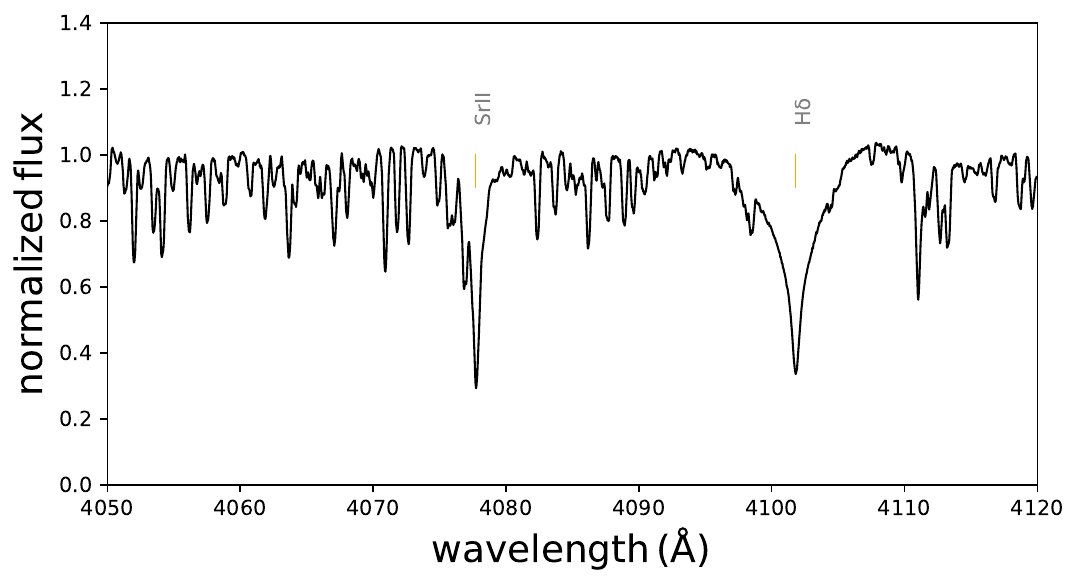}
\vskip 0.25cm
\caption{Spectrum of the star HD\,72968 in the range of 4050 \AA{} to 4120 \AA{}, showing chemical elements such as Strontium Sr\,II at 4077.7 \AA{} and H\,$\delta$ at 4101.7 \AA{}.}
\label{fig:Fig. 2.5}
\end{figure}

In Figure \ref{fig:Fig. 2.5}, two significant absorption lines corresponding to Sr\,II and H\,$\delta$ are observed, as established by \citet{1958ApJ...128..228B} for HD\,72968 of spectral type A2p, where the predominant chemical element is Sr (Strontium). \citet{1931ApJ....73..104M} defined that, in general, stars exhibit approximately equal chemical abundances. However, some stars do not fit within the bidimensional temperature and pressure scale, such as those displaying strong Sr (Strontium) lines at 4077 \AA{}. According to the Henry Draper catalog, these are classified as chemically peculiar stars. Additionally, the Balmer line H\,$\delta$ at 4101 \AA{} is present, which is characteristic of spectral type A stars, such as HD\,72968.

\begin{figure}[!h]
\centering
\includegraphics[width=0.8\textwidth]{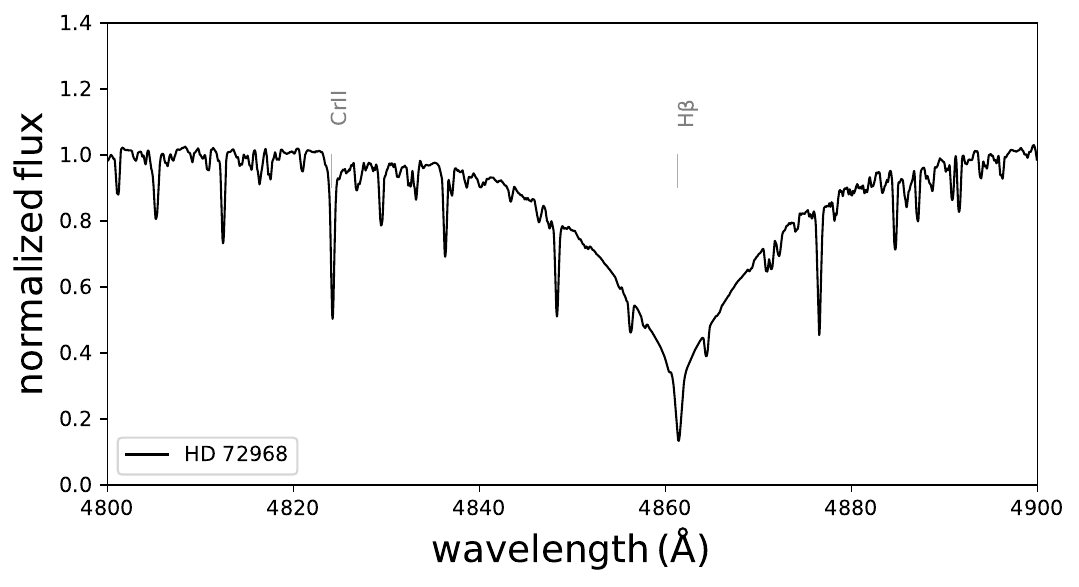}
\vskip 0.25cm
\caption{Spectrum of the star HD\,72968 in the range of 4800 \AA{} to 4900 \AA{}, showing chemical elements such as Chromium II at 4824.1 \AA{} and H\,$\beta$ at 4861.3 \AA{}.}
\label{fig:Fig. 2.6}
\end{figure}

In Figure \ref{fig:Fig. 2.6}, the spectrum of HD\,72968 is presented in the range of 4800 \AA{} to 4900 \AA{}, where two prominent absorption lines are identified as Cr\,II at 4824.1 \AA{} and H\,$\beta$ at 4861.3 \AA{}. The Sr\,II and Cr\,II elements present in HD\,72968 correspond to the Sr-Cr spectral type classification for this star, as determined by \citet{1974A&A....37..367H}.

\subsubsection{Chemical abundance}\label{Sec:Sec. 3.0.1}
We have performed the chemical abundance analysis presented in Table \ref{Tab: Tab. 2.9} for the star HD\,72968 in the wavelength range from 4800 \AA{} to 5000 \AA{} using the \texttt{IRAF} software. The wavelengths of each absorption line, equivalent widths, and errors were obtained. Then, through the astronomical catalog VizieR\footnote{\url{http://vizier.cds.unistra.fr/}}, the chemical elements present were identified based on the peculiar chemical characteristics of HD\,72968.

Using the previously compiled data from the NIST Atomic Spectra Database Lines Data\footnote{\url{https://physics.nist.gov/}}, we obtained the parameters $E_{\mathrm{low}}$ (cm$^{-1}$), $E_{\mathrm{high}}$ (cm$^{-1}$), $g_k A_{ki}$ ($10^8$ s$^{-1}$), $f_{ik}$, and $\log(g_i f_{ik})$, which are essential for the chemical abundance analysis of the chemically peculiar star.

To derive the abundance values from the parameters retrieved from NIST, the software ABUNDANCE SPECTRUM was used. The stellar parameters of HD\,72968, including its rotational velocity, were input to ensure a more accurate determination of the star's chemical abundances.

The chemical abundance values [X/H] of each element relative to the solar abundance are presented in Table~\ref{Tab: Tab. 2.10}.

\begin{table}[]
\caption{Equivalent width of the most representative spectral lines of HD\,72968.}
\centering
\begin{tabular}{l c c c c c }
\hline
\textrm{Elements} & \textrm{Wavelength (\AA)} & \textrm{Z} & \textrm{Equi. Width (\AA)} & \textrm{error (\AA)} \\
\hline
\hline
\noalign{\smallskip}
 Ti\,II   & 4805.1050 & 22 & 0.1192 & 0.01271 \\
 Ti\,I    & 4812.2400 & 22 & 0.1206 & 0.01182 \\
 Cr\,II   & 4824.1300 & 24 & 0.2590 & 0.01900 \\
 Ni\,I    & 4829.6800 & 28 & 0.1011 & 0.01289 \\
 Cr\,II   & 4836.1250 & 24 & 0.1261 & 0.03728 \\
 Cr\,II   & 4848.4100 & 24 & 0.1481 & 0.00625 \\
 H$\beta$ & 4861.3320 &  1 & 11.720 & 0.06000 \\
 Cr\,II   & 4876.4100 & 24 & 0.2038 & 0.00629 \\
 Cr\,II   & 4884.5700 & 24 & 0.1014 & 0.01266 \\
 Fe\,I    & 4891.5500 & 26 & 0.0718 & 0.01080 \\
 S\,II    & 4901.6500 & 27 & 0.1049 & 0.02310 \\
 Ni\,I    & 4912.4900 & 28 & 0.0812 & 0.01145 \\
 Nd\,II   & 4920.2800 & 60 & 0.1717 & 0.03485 \\
 Fe\,II   & 4923.9210 & 26 & 0.1776 & 0.04333 \\
 Fe\,I    & 4930.3310 & 26 & 0.0444 & 0.01231 \\
 Cr\,I    & 4936.1550 & 24 & 0.0539 & 0.01418 \\
 Ti\,I    & 4941.0150 & 22 & 0.0512 & 0.01143 \\
 Ba\,II   & 4957.1500 & 56 & 0.1981 & 0.01404 \\
 Sr\,I    & 4971.6680 & 38 & 0.0481 & 0.00947 \\
 Ti\,I    & 4977.7310 & 22 & 0.1447 & 0.02377 \\
 Y\,II    & 4982.1300 & 39 & 0.0374 & 0.00881 \\
\hline
\end{tabular}
\label{Tab: Tab. 2.9}
\end{table}

\begin{table}[]
\caption{Atomic parameters and chemical abundances for HD\,72968}
\centering
\begin{tabular}{lccrccrcccc}
\hline
\noalign{\smallskip}
Elements & $\lambda$ (\AA) & Z & $E_{\mathrm{low}}$ (cm$^{-1}$) & $E_{\mathrm{high}}$ (cm$^{-1}$) & $g_k A_{ki}$ ($10^8$ s$^{-1}$) & $f_{ik}$ & $\log(g_i f_{ik})$ & [X/H] & $\sigma$ \\
\hline
\hline
\noalign{\smallskip}
Ti\,II  & 4805.0928 & 22 &  16625.2441 &  37430.681  & 2.200e7   & 1.900E-2 & -1.120   & -0.387  & 0.179 \\
Ti\,I   & 4812.8934 & 22 &   6842.9620 &  27614.679  & 9.900e4   & 3.100E-5 & -3.460   &  2.678  & 0.179 \\
Cr\,II  & 4824.1308 & 24 &  31219.3322 &  51942.664  & 1.700e7   & 5.900E-3 & -1.230   &  0.703  & 0.179 \\
Ni\,I   & 4829.0230 & 28 &  28569.2030 &  49271.540  & 1.300e8   & 9.300E-2 & -0.330   & -0.557  & 0.220 \\
Cr\,II  & 4836.2295 & 24 &  31117.3254 &  51788.815  & 1.600e6   & 9.400E-4 & -2.250   &  0.990  & 0.182 \\
Cr\,II  & 4848.2497 & 24 &  31168.5755 &  51788.815  & 2.100e7   & 9.200E-3 & -1.130   & -0.001  & 0.179 \\
H$\beta$& 4861.3330 & 1  &  82259.1580 & 102823.904  & 2.694e8   & 1.194E-1 & -0.020   &  4.092  & 0.188 \\
Cr\,II  & 4876.4892 & 24 &  31168.5755 &  51669.406  & 9.600e6   & 4.300E-3 & -1.470   &  0.636  & 0.179 \\
Cr\,II  & 4884.6035 & 24 &  31117.3254 &  51584.100  & 2.300e6   & 1.400E-3 & -2.080   &  0.668  & 0.179 \\
Fe\,I   & 4891.4920 & 26 &  22996.6740 &  43434.627  & 2.160e8   & 8.590E-2 & -0.112   & -2.597  & 0.179 \\
S\,II   & 4901.2780 & 27 & 130134.1580 & 150531.304  & 9.600e7   & 9.600E7  & -0.460   & 10.831  & 0.180 \\
Ni\,I   & 4912.0180 & 28 &  30392.0030 &  50744.552  & 4.500e7   & 5.400E-2 & -0.790   & -0.014  & 0.179 \\
Nd\,II  & 4920.6850 & 60 &    513.3300 &  20830.030  & 7.100e7   & 2.580E-2 & -0.588   &  1.270  & 0.182 \\
Fe\,II  & 4923.9212 & 26 &  23317.6351 &  43620.984  & 1.700e7   & 1.040E-2 & -1.210   & -1.745  & 0.184 \\
Fe\,I   & 4930.3149 & 26 &  31937.3250 &  52214.345  & 1.720e7   & 2.100E-2 & -1.201   & -0.690  & 0.179 \\
Cr\,I   & 4936.3357 & 24 &  25106.2953 &  45358.584  & 1.300e8   & 6.600E-2 & -0.340   & -0.125  & 0.179 \\
Ti\,I   & 4941.5707 & 22 &  17423.8560 &  37654.690  & 2.600e7   & 1.900E-2 & -1.010   &  1.030  & 0.179 \\
Ba\,II  & 4957.0950 & 56 &  48258.6170 &  68426.095  & 4.100e8   & 2.500E-1 &  0.190   &  4.162  & 0.179 \\
Sr\,I   & 4971.6680 & 38 &  14898.5450 &  35006.908  & 3.900e6   & 2.900E-3 & -1.800   &  3.239  & 0.179 \\
Ti\,I   & 4977.7275 & 22 &  16267.4800 &  36351.365  & 4.850e7   & 1.640E-2 & -0.740   &  1.216  & 0.180 \\
Y\,II   & 4982.1300 & 39 &   8328.0390 &  28394.177  & 1.380e7   & 7.300E-3 & -1.290   &  0.408  & 0.179 \\
\hline
\end{tabular}
\label{Tab: Tab. 2.10}
\end{table}

\ \\
The uncertainty in the chemical abundances, $\sigma$, was determined based on the error propagation of several parameters obtained in this work, including the uncertainty in the effective temperature ($E_{T_{\mathrm{eff}}}$), surface gravity ($E_{logg}$), rotational velocity ($E_{v \sin i}$), and the equivalent width uncertainty derived from Gaussian fitting in IRAF ($E_{\mathrm{eqw}}$), using equation \ref{eq: eq. 2.18}.

\begin{equation}
\mathrm{\sigma = \sqrt{(\frac{E_{T_{eff}}} {T_{eff}})^2+(\frac{E_{logg}} {logg})^2+(\frac{E_{v \sin i}} {v \sin i})^2+({E_{\mathrm{eqw}}})^2}},
\label{eq: eq. 2.18}
\end{equation}
\vskip 0.25cm

\begin{figure}[h]
\centering
\includegraphics[width=0.6\textwidth,angle=0]{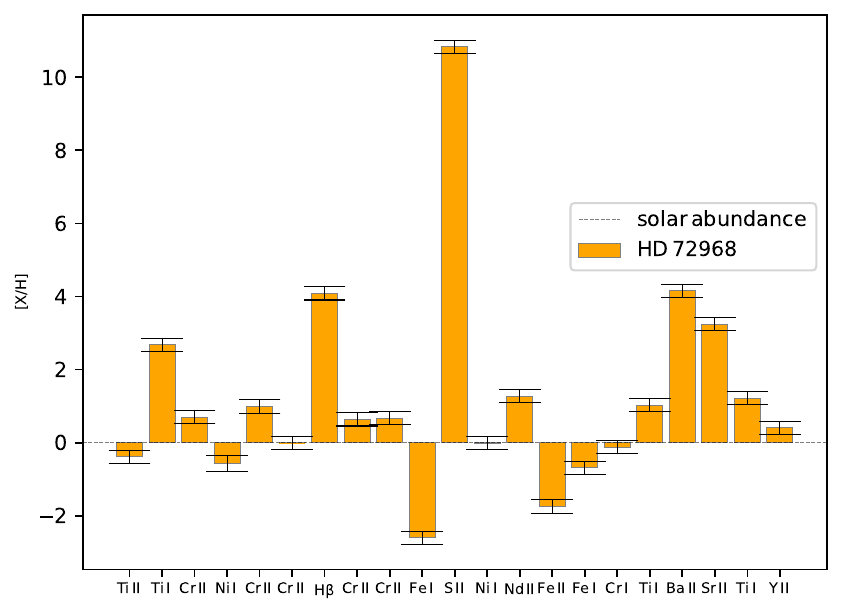}
\vskip 0.25cm
\caption{Comparison of the chemical abundances of the star HD~72968 with those of the Sun. Positive values indicate an overabundance of certain elements in the star, while negative values reflect a deficiency. Notable differences in elements such as barium (Ba), sulfur (S), and strontium (Sr) are characteristic of chemically peculiar stars.}
\label{fig:Fig. 2.14}
\end{figure}
\ \\
\ \\
\subsubsection{Numerical method}\label{Sec:Sec. 3.0.2}

Obtained the spectrum corrected for the Doppler effect, the search for the star's data in the Gaia DR3 Mission database\footnote{\url{https://gea.esac.esa.int/archive/}} was carried out to obtain the metallicity [M/H] in order to download the stellar atmospheres from ATLAS9. The acquisition of the ATLAS9 models\footnote{\url{https://wwwuser.oats.inaf.it/castelli/grids.html}} was performed using the code "am10ak2c125odfnew" \citep{2003IAUS..210P.A20C}.

In the present method, the chi-square statistical model was used to obtain the best fit between the observed spectrum and the theoretical spectrum obtained from the SPECTRUM software with stellar atmosphere models from ATLAS9\footnote{\url{https://wwwuser.oats.inaf.it/castelli/grids.html}}.

\ \\

By estimating different stellar models according to temperatures in the SPECTRUM software, it was determined that the best chi-square value was estimated between the effective temperatures of $\mathrm{T_{eff}}$ = 7000 K to 15000 K.

For the elaboration of the table, the wavelength range from 4800 \AA{} to 4900 \AA{} was used, defining the effective temperatures of $\mathrm{T_{eff}}$ = 7000 K to 15000 K in steps of 250 K. For the surface gravity values, they were established from logg = 0.0 dex to 5.0 dex in steps of 0.5 dex, the microturbulence velocity set at 0.0 $\mathrm{km\,s^{-1}}$ according to ATLAS9 stellar models, solar metallicity value defined as [M/H] = 0.00, and spectral resolution of r = 0.01483899664.

The mixing length parameter, which corresponds to the disruption and dispersion of bubbles traveling in a convective fluid over a distance \textquotedblleft{l}\textquotedblright from their equilibrium position in a scale atmosphere \textquotedblleft{H}\textquotedblright \citep{2023A&A...670A..94R}, was set at 1.25 l/H according to the ATLAS9 stellar models. The rotational velocity value was established based on the references of \citet{1971ApJ...164..309P}, defined as 16 $\mathrm{km\,s^{-1}}$, determined through the extended helium abundance in the stellar atmosphere $\mathrm{H_e}$(ext) = 0.7 kilogauss. 

The rotational velocity value according to \citet{1974A&A....37..367H} was in the range of 10 to 12 $\mathrm{km\,s^{-1}}$, following the criterion of weak magnetic lines of specific chemical elements such as Fe I and Cr I, as well as the common rotational broadening in all lines of HD\,72968. Knowing the rotational velocity data, a range from 0 to 100 $\mathrm{km\,s^{-1}}$ in steps of 1 $\mathrm{km\,s^{-1}}$ was established to determine which value best fit the chi-square in the theoretical spectrum. Evaluating the aforementioned parameters, the best model converged to the following value $\mathrm{v_{sini} = 8 \mathrm{km\,s^{-1}}}$, which showed the best fit determined through the chi-square statistical method.

\ \\

The ATLAS9 grid\footnote{\url{https://wwwuser.oats.inaf.it/fiorella.castelli/grids.html}}, consisting of effective temperature versus surface gravity, was constructed, obtaining the best chi-square with the parameters shown in Table \ref{Tab: Tab. 2.5}.

\vskip 0.25cm
\begin{table}[]
\caption{Parameters established through the mathematical method for the best chi-square fit in the theoretical spectrum for HD\,72968, which include: effective temperature, surface gravity, microturbulence velocity, rotational velocity, mixing length, metallicity, best chi-square fit, and spectral resolution.}
\centering
\resizebox{0.4\textwidth}{3.1cm}{
\begin{tabular}{l l}
\hline
\textrm{Parameters} & \textrm{Values} \\
\hline
\hline
$\rm{T_{eff}}$          & $\rm{10250 \pm 250\,K} $           \\
$\rm{log{g}}$           & $\rm{4.0  \pm 0.5\,dex} $          \\
$\rm{*V_{micro}}$       & $\rm{0.0 \,{km\,s^{-1}}}$          \\
$\rm{v\sin{i}}$         & $\rm {8 \pm 1 \, {km\,s^{-1}}}$    \\
$\rm{[l/H]}$            & $\rm{1.25}$                        \\
$\rm{[M/H]}$            & $\rm{0.00}$                        \\
$\rm{\chi^2}$           & $\rm{18.5}$                        \\
$\rm{r}$           & $\rm{0.01483899664}$                    \\
\hline
\end{tabular}
}
\label{Tab: Tab. 2.5}
\end{table}

\begin{figure}[!ht]
\centering
\includegraphics[trim=0.6cm 0.6cm 0.6cm 0.6cm, width=1\textwidth,angle=0]{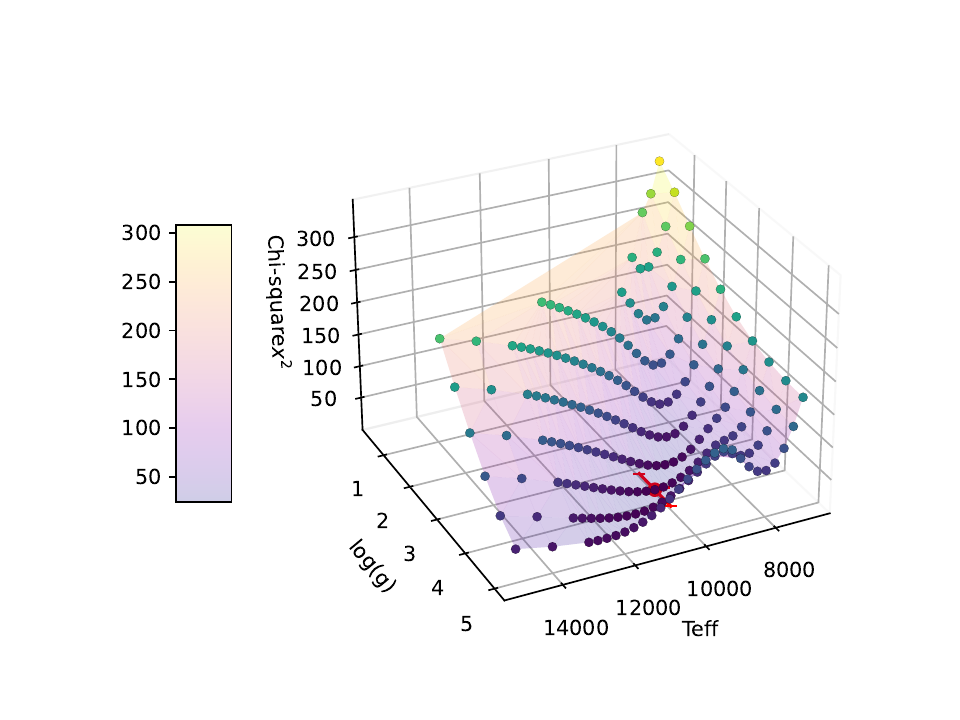}
\vskip 0.25cm
\caption{Chi-square analysis to determine the best normalized spectrum in reference to the observed spectrum of HD\,72968. The $\mathrm{\chi^2}$ graph was created using only 3 degrees of freedom, showing the minimum chi-square value of $\mathrm{\chi^2}$ = 18.5, represented by a red point. Each point within the graph represents an effective temperature associated with its surface gravity.}
\label{fig:Fig. 2.7}
\end{figure}

\begin{figure}[!ht]
\centering
\includegraphics[width=0.8\textwidth]{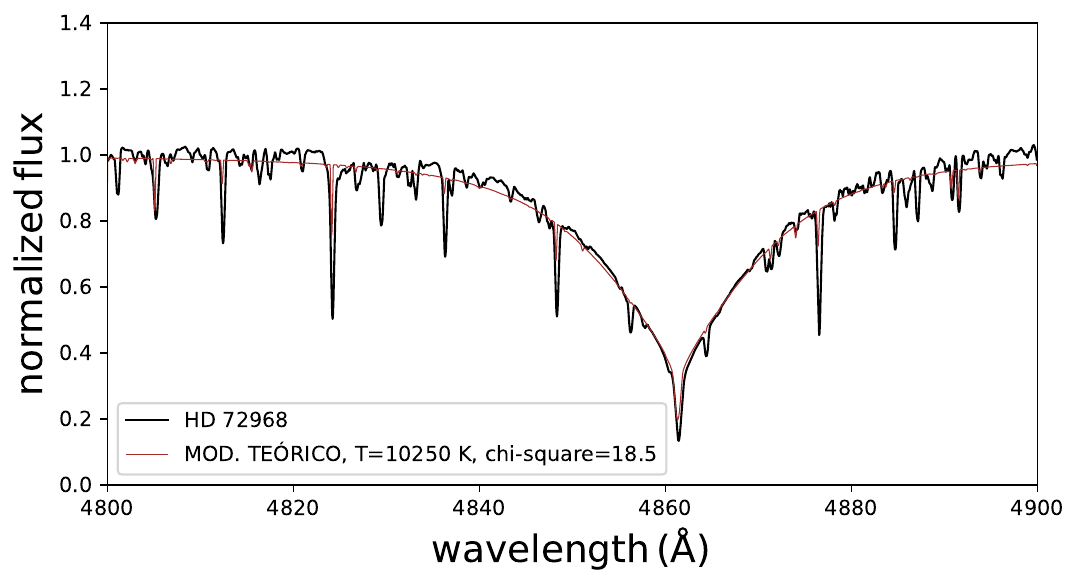}
\vskip 0.25cm
\caption{Comparison between the observed spectrum (Black) of CP HD\,72968 and the theoretical spectrum (Red) with the smallest standard deviation, from 4800.0 \AA{} to 4900.0 \AA{}. The obtained temperature is $\mathrm{T_{eff} = 10250\, K}$, and $\mathrm{\log g = 4.0\,dex}$.}
\label{fig:Fig. 2.8}
\end{figure}

In Figure \ref{fig:Fig. 2.8}, it can be seen that the theoretical spectrum from the SPECTRUM software, starting at 4800.0 \AA{}, ending at 4900 \AA{}, and having a resolution of r = 0.01483899664, exhibits spectral similarities in the Balmer line and in the absorption lines of the observed spectrum. It is important to highlight that modeling stellar atmospheres with ATLAS9 and using SPECTRUM to generate the graph serve as guidance and provide direction for the research, but are not meant to be taken as absolute results. "Models like ATLAS9 and MARCS serve as starting points for our technique, which is by no means limited to them" \citep{2023A&A...675A.191W}.

\subsubsection{Comparative method}\label{Sec:Sec. 3.0.3}

To determine thermal criteria, stars of the same luminosity class corresponding to different spectral types are compared whenever possible. However, since the temperature sequence covers such a wide range from O to M,it is logical to understand that these criteria gradually vary within a certain range where they respond perfectly to some criterion. It often happens that upon reaching a certain spectral type, the difference between the intensities of the compared lines reaches a proportion so large that it can no longer be applied as an indicator \citep{1968BAAA...13...13J}.

The theoretical effective temperature $\mathrm{T_{eff} = 10250 \pm 250 \,K}$ was obtained through spectral analysis using the Spectrum software. A comparative method with other peculiar stars was then applied to define the upper and lower limits for the star HD\,72968. For this comparison method, data from two chemically peculiar stars with similar characteristics to the studied star HD\,72968 were extracted from the SIMBAD database\footnote{\url{https://simbad.cds.unistra.fr/simbad/}} using a special search type from the Hipparcos catalog, which has been identifying bright stars with precise and independent information for the past thirty years \citep{2021ApJS..254...42B}. The selected peculiar A-type stars were HD\,220825 and HD\,223640.

HD\,220825 is cataloged as a chemically peculiar star of spectral type Ap, with an effective temperature of 9470 K, a surface gravity of logg = 4.20 dex, and a stellar radius of 1.71 $\mathrm{R_{\odot}}$, determined through spectroscopic methods according to \citet{2021A&A...655A.106R}. HD\,223640 is a chemically peculiar star of spectral type Ap, with an effective temperature of 12340 K, determined photometrically by \citet{1985A&A...144..191L} using the Blackwell-Shallis method, one of the most precise methods for determining a star's effective temperature through photometry, developed by \citet{1977MNRAS.180..177B}. Additionally, the temperature value of 12340 K was confirmed through the empirical method for determining effective temperatures and angular diameters in milliarcseconds, established by \citet{1988A&AS...72..551M}.

To compare the spectral characteristics, two spectra were used, as presented in Table \ref{Tab: Tab. 2.6}. The spectra used in this research fall within the blue arm, corresponding to wavelengths from 300 nm to 500 nm, according to the European Southern Observatory (ESO) database\footnote{\url{https://www.eso.org/public/}}, where ultraviolet light is used for various astrophysical studies. These spectra are cataloged in scientific spectra, and diffraction gratings of 437 l/mm (lines per millimeter) were used for all spectra.

\vskip 0.25cm
\begin{table}[]
\caption{Summary of UVES spectra extracted for comparison with the star HD\,72968. N represents the number of extracted spectra, HJD is the Heliocentric Julian Date, S/N corresponds to the signal-to-noise ratio, and R is the instrument resolution.}
\centering
\begin{tabular}{l c c c c c c c}
\hline
Object      & Date         & HJD           & N     & Exposure Time (s)     & S/N       & R\\
\hline
\hline
HD 220825   & 17-05-2016    & 2457525.91    & 1     & 28.0017                   & 149.7     & 80 000\\
HD 223640   & 14-08-2002    & 2452500.94    & 1     & 40.0009                   & 151.33    & 80 000\\
\hline
\end{tabular}
\label{Tab: Tab. 2.6}
\end{table}

\begin{figure}[h]
\centering
\includegraphics[trim=0.1cm 0.0cm 0.1cm 0.1cm,clip,width=0.8\textwidth,angle=0]{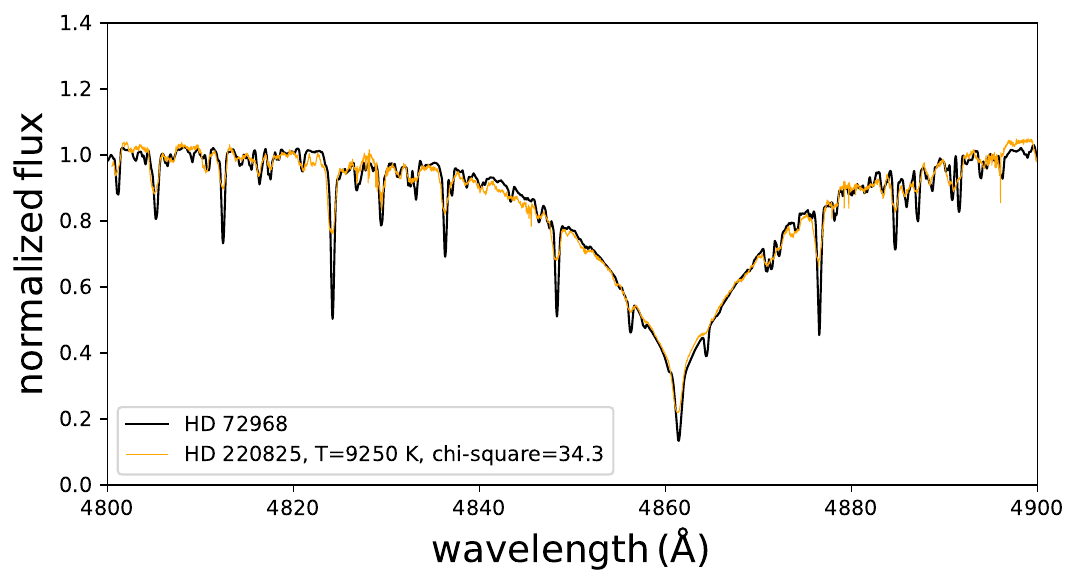}
\vskip 0.25cm
\caption{Comparison of the chemically peculiar star HD\,220825 (orange) with 9250 K against the spectrum of the chemically peculiar star HD\,72968 (black).}
\label{fig:Fig. 2.9}
\end{figure}

The spectra of the stars HD\,220825 and HD\,223640 were acquired from the European Southern Observatory (ESO) database\footnote{\url{https://www.eso.org/public/}}, normalized, and wavelength\textrm{-}calibrated for subsequent comparison with the star HD\,72968 to estimate the effective temperature. Figure \ref{fig:Fig. 2.9} shows the comparison between two chemically peculiar stars with similar spectral types. The spectrum of the star HD\,220825 is shown in orange, while the spectrum of the star HD\,72968 is shown in black. The chi-square value obtained between both stars was $\mathrm{\chi^2 = 34.3}$, which is greater than the theoretical chi-square value, leading us to consider the star HD\,220825 as the reference for the lower effective temperature due to its spectral visual similarity.

\begin{figure}[!ht]
\centering
\includegraphics[trim=0.1cm 0.0cm 0.1cm 0.1cm,clip,width=0.8\textwidth,angle=0]{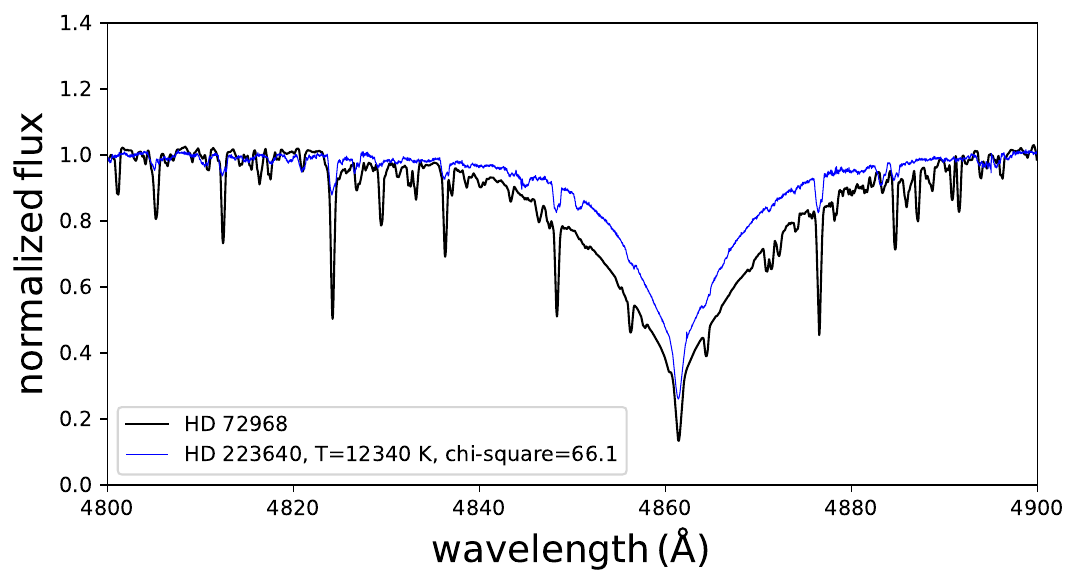}
\vskip 0.25cm
\caption{Comparison of the chemically peculiar star HD\,223640 (blue) with a correction factor relative to the chemically peculiar star HD\,72968 (black).}
\label{fig:Fig. 2.10}
\end{figure}

Figure \ref{fig:Fig. 2.10} shows the comparison between two chemically peculiar stars with similar spectral types. The spectrum of the star HD\,223640 is shown in blue, while the spectrum of the star HD\,72968 is shown in black. The chi-square value obtained between both stars was $\mathrm{\chi^2} = 66.1$, which is greater than the theoretical chi-square value, leading us to consider the star HD\,223640 as the reference for the upper effective temperature. Since the effective temperature is higher than the theoretical effective temperature of 10250 K, the chi-square value is also higher. The blue spectrum near the wavelength 4861.3 \textrm{\AA} presents spectral disparity compared to the black spectrum.

\begin{figure}[]
\centering
\includegraphics[trim=0.1cm 0.0cm 0.1cm 0.1cm,clip,width=0.8\textwidth,angle=0]{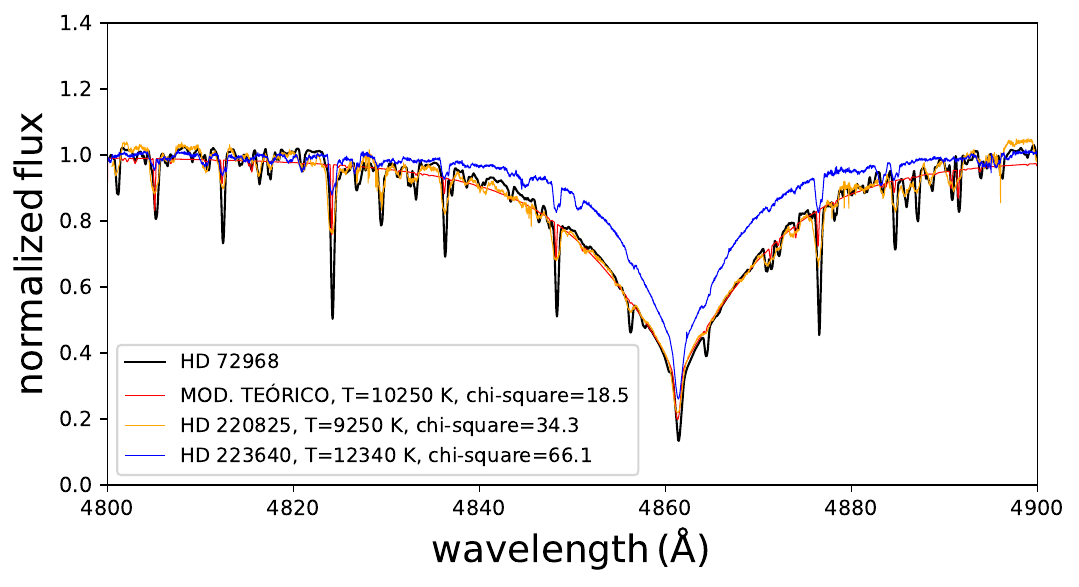}
\vskip 0.25cm
\caption{Spectral comparison of the star HD\,72968 (black) with respect to the theoretical spectrum (red), the star HD\,220825 (orange), and the star HD\,223640 (blue).}
\label{fig:Fig. 2.11}
\end{figure}

Figure \ref{fig:Fig. 2.11} shows the complete comparison of the spectra used for the spectral comparison with the chi-square statistical technique. The black spectrum corresponds to the observed spectrum of HD\,72968, the red spectrum belongs to the theoretical spectrum for the star HD\,72968, the orange spectrum corresponds to the peculiar star HD\,220825, and the blue spectrum belongs to the star HD\,223640.

\section{Spectral distribution of energy SED}\label{Sec:Sec. 4}

The spectral energy distribution (SED) is a graphical representation or a curve that relates the flux emitted by a star in units of Flux $\mathrm{(erg s^{-1}cm^{-2}\AA{}^{-1})}$ and the wavelength in units of Angstrom (\AA{}). Within the SED, two variables are presented: the photometric points emitted by the star, in this case, HD\,72968, and a continuous spectrum ranging from the ultraviolet to the far-infrared, which corresponds to the stellar model of the best chi-square fit between the variables.

In Table \ref{Tab: Tab. 2.7}, the data of fluxes and photometric flux errors, wavelengths, and the information of the filters used for the observation of the star HD\,72968 are indicated, extracted from SED VizieR\footnote{\url{http://vizier.cds.unistra.fr/vizier/sed/}}.

To define the SED, the Kurucz ODFNEW / NOVER (2003) model \citet{2003IAUS..210P.A20C} was selected, as this model encompasses stars of different spectral types, including stars with chemical abundances. Additionally, it presents improved opacities and abundances in the energy spectra. The values obtained through the spectroscopic study were then input: effective temperature $\mathrm{10000 \,a\, 10500 \,K}$, surface gravity logg $\mathrm{3.5 \, \,a \, \, 4.5 \, dex}$, solar metallicity of 0.0, and relative abundance of alpha elements of 0.0, simulating the solar chemical abundance. Furthermore, we considered the reddening caused by extinction in the line of sight due to interstellar dust, such as $\mathrm{E(B-V)_{SFD}}$ and $\rm{Av_{SFD}}$, which were calculated using equation \ref{eq: eq. 2.4} through IRSA\footnote{\url{https://irsa.ipac.caltech.edu/}}.

\begin{equation}
\mathrm{Av_{SFD} / E(B-V)_{SFD}= 0.31},
\label{eq: eq. 2.4}
\end{equation}
\vskip 0.25cm

\ \\
\noindent
the best fit within the range of free parameters that we set based on previous results leads to a convergence of effective temperature $\mathrm{T_{eff}=10000\,K}$, a surface gravity of logg = 4.5 dex, metallicity = 0.5, and a visual extinction of Av = 0.195. 

It is important to highlight the variation in metallicity from 0.0 to the resulting metallicity of 0.5, which is higher than solar. This suggests that the star HD\,72968 exhibits chemical abundance anomalies or another effect, such as rapid rotation, which could increase the metallicity value of HD\,72968 relative to solar metallicity \citet{2023A&A...675A..54V}.

Obtaining the photometric parameters through the first construction of the SED, we have used equation \ref{eq: eq. 2.7} obtained from \citet{1997A&AS..122...51K}, which relates the initial metallicity $\mathrm{[M/H]_{0}}$ and the effective temperature $\mathrm{T_{eff}}$ to estimate the final metallicity [M/H] of HD\,72968.

\begin{equation}
\mathrm{[M/H]=0.260+1.193 \cdot [M/H]_{0} \mathrm{-}3.4e-5 \cdot \mathrm{T_{eff}}},
\label{eq: eq. 2.7}
\end{equation}
\vskip 0.25cm

\ \\
\noindent
where $\rm{M/H}$ is the actual stellar metallicity of the chemically peculiar star estimated to be $\rm{M/H=0.517 \pm 0.008}$, which was adjusted to the approximate value of the effective temperature determined through spectroscopy, $\mathrm{{M/H_{0}}}$ is the estimated initial stellar metallicity $\mathrm{M/H_{0} = 0.5}$ according to the SED, and $\mathrm{T_{eff}}$ is the effective temperature of the star estimated to be $\mathrm{T_{eff} = \rm{10000 \pm 250\,K} }$. With the values obtained from the SED, it was verified that the star HD\,72968 exhibits chemical abundances with a value of $\mathrm{M/H > 0}$, which is the solar metallicity reference for all stars.

\vskip 0.25cm
\begin{table}[]
\caption{Photometric summary for the mCP star HD\,72968 in different bandwidths obtained from the SED of VizieR.}
\normalsize
\begin{center}
\begin{tabular}{lccrc}
\hline
\noalign{\smallskip}
\textrm{wavelength} & \textrm{Flux} & \textrm{Flux error} & \textrm{Filter}\\
\textrm{$(\mu m)$} & \textrm{$(erg s^{-1}cm^{-2}\mu m^{-1})$} & \textrm{$(erg s^{-1}cm^{-2}\mu m^{-1})$} & \textrm{}\\
\hline
\hline
\textrm{0.4203} & \textrm{3.46e-7} & \textrm{7.13e-13 } & 
\textrm{HIP:BT}\\
\textrm{0.4442} & \textrm{3.40e-7} & \textrm{1.35e-12 } & 
\textrm{Johnson:B}\\
\textrm{0.5036} & \textrm{2.21e-7} & \textrm{5.95e-13 } & 
\textrm{GAIA3:Gbp}\\
\textrm{0.5319} & \textrm{2.06e-7} & \textrm{5.64e-13 } & 
\textrm{HIP:VT}\\
\textrm{0.5319} & \textrm{2.08e-7} & \textrm{1.13e-12 } & 
\textrm{HIP:VT}\\
\textrm{0.613} & \textrm{7.19e-8} & \textrm{4.89e-14 } & 
\textrm{PAN-STARRS/PS1:r}\\
\textrm{0.762} & \textrm{6.97e-8} & \textrm{3.93e-13 } & 
\textrm{GAIA3:Grp}\\
\textrm{0.7485} & \textrm{4.40e-8} & \textrm{4.01e-14 } & 
\textrm{PAN-STARRS/PS1:i}\\
\textrm{0.7635} & \textrm{4.23e-8} & \textrm{3.93e-14 } & 
\textrm{SDSS:i}\\
\textrm{1.239} & \textrm{1.61e-8} & \textrm{5.08e-13 } & 
\textrm{2MASS:J}\\
\textrm{1.25} & \textrm{1.61e-8} & \textrm{3.60e-13 } & 
\textrm{Johnson:J}\\
\textrm{1.63} & \textrm{5.86e-9} & \textrm{3.49e-13 } & 
\textrm{Johnson:H}\\
\textrm{1.65} & \textrm{5.77e-9} & \textrm{3.45e-13 } & 
\textrm{2MASS:H}\\
\textrm{2.164} & \textrm{2.29e-9} & \textrm{8.31e-14 } & 
\textrm{2MASS:Ks}\\
\textrm{2.19} & \textrm{2.16e-9} & \textrm{8.21e-14 } & 
\textrm{Johnson:K}\\
\textrm{3.35} & \textrm{4.59e-10} & \textrm{4.47e-14 } & 
\textrm{WISE:W1}\\
\textrm{3.35} & \textrm{4.30e-10} & \textrm{8.05e-14 } & 
\textrm{WISE:W1}\\
\textrm{3.4} & \textrm{3.84e-10} & \textrm{1.76e-13 } & 
\textrm{Johnson:L}\\
\textrm{4.6} & \textrm{1.38e-10} & \textrm{1.63e-14 } & 
\textrm{WISE:W2}\\
\textrm{4.6} & \textrm{1.16e-10} & \textrm{7.15e-15 } & 
\textrm{WISE:W2}\\
\textrm{5.03} & \textrm{1.04e-10} & \textrm{2.56e-14 } & 
\textrm{Johnson:M}\\
\textrm{8.61} & \textrm{1.26e-11} & \textrm{3.48e-15 } & 
\textrm{AKARI:S9W}\\
\textrm{11.56} & \textrm{3.10e-12} & \textrm{5.19e-16 } & 
\textrm{WISE:W3}\\
\textrm{22.09} & \textrm{2.48e-13} & \textrm{2.04e-16 } & 
\textrm{WISE:W4}\\
\hline
\end{tabular}
\end{center}
\label{Tab: Tab. 2.7}
\end{table}

Finally, for better accuracy, we developed a theoretical spectral energy distribution (SED) to obtain the best fit. The statistical technique of chi-square was applied, using the total theoretical spectral flux from \citet{2005AJ....130.1127F} with equation \ref{eq: eq. 2.8}, complemented by equations \ref{eq: eq. 2.9} and \ref{eq: eq. 2.10} obtained from \citet{2018MNRAS.476.3039R} for the determination of the partial stellar flux and the normalized extinction curve, respectively, over a wavelength range from 1005 \AA{} to 1600000 \AA{} covering from ultraviolet to infrared.

For the total theoretical spectral flux, the photometric data offered by the VizieR SED\footnote{\url{http://vizier.cds.unistra.fr/vizier/sed/}} and theoretical photometric models gathered from VOSA\footnote{\url{http://svo2.cab.inta-csic.es/theory/newov2/index.php?models=Kurucz2003all}} were compared, with effective temperature parameters ranging from 10000 K to 10500 K in steps of 250 K, surface gravity from 3.5 to 4.5 dex in steps of 0.5 dex, and solar

\begin{equation}
\mathrm{f\lambda=f_{\lambda,0} 10^{-0.4E(B-V)[k(\lambda-V)+R(V)]}},
\label{eq: eq. 2.8}
\end{equation}
\vskip 0.25cm

where,

\begin{equation}
\mathrm{f_{\lambda,0}=\left(\frac{R_1}{d}\right)^2 \cdot f_{\lambda,1}},
\label{eq: eq. 2.9}
\end{equation}
\vskip 0.25cm

\ \\
\noindent
$\mathrm{E(B-V) = 0.0380 \pm 0.0006}$ corresponds to the data obtained from IRSA\footnote{url{https://irsa.ipac.caltech.edu/}}, where $\mathrm{R(V) = A_v / E(B-V)}$. The value of $\mathrm{k(\lambda - V) = E(\lambda - V)/E(B - V)}$ was determined using equation \ref{eq: eq. 2.10}.

where $\mathrm{R_{1}}$ is the radius to be determined, adjusted from $\mathrm{2.00 \,to\, 2.10 \,R_{\odot}}$, corresponding to the radius of the star HD\,72968, $\mathrm{d=107.531 \pm 1.135\,pc}$ is the photometric distance obtained from the Gaia DR3 mission\footnote{\url{https://gea.esac.esa.int/archive/}}, and $\mathrm{f_{\lambda,1}}$ is the theoretical flux of the star related to the wavelength.

\begin{equation}
\mathrm{k =} \left\lbrace
\begin{array}{ll}
\mathrm{\epsilon \lambda^{-\beta} - R_v} \,\,\,\,\,\,\,\,\,\,\,\,\,\,\,\,\,\,\,\,\,\,\,\,\,\,\,\,\,\,\,\textup{if } x<0.3\\
\mathrm{R_v \left(a(x) + \frac{b(x)}{R_v} - 1\right) \,\,\,\,\,\textup{if } \leq 0.3 \,x \leq 8.0},
\end{array}
\right.
\label{eq: eq. 2.10}
\end{equation}
\vskip 0.25cm

\ \\
\noindent
where $\mathrm{x=1/\lambda\,\mu m^{-1}}$, the color excess ratios with B, V normalization include $\mathrm{\epsilon = 1.19}$, $\mathrm{\beta = 1.84}$, $\mathrm{R_v=3.05}$ obtained from \citet{1990ApJ...357..113M}, $\mathrm{a(x)}$ and $\mathrm{b(x)}$ are determined by the equations \ref{eq: eq. 2.11} and \ref{eq: eq. 2.12} for values in the range of $\mathrm{0.3\,\mu m^{-1}}$ to $\mathrm{1.1\,\mu m^{-1}}$, the equations \ref{eq: eq. 2.13} and \ref{eq: eq. 2.14} are used for values in the range of $\mathrm{1.1\,\mu m^{-1}}$ to $\mathrm{3.3\,\mu m^{-1}}$, including $\mathrm{y=x-1.82}$ as expressed by \citet{1989ApJ...345..245C}.

\begin{equation}
\mathrm{a(x)= 0.574x^{1.61}},
\label{eq: eq. 2.11}
\end{equation}
\vskip 0.25cm

\begin{equation}
\mathrm{b(x)= -0.527x^{1.61}},
\label{eq: eq. 2.12}
\end{equation}
\vskip 0.25cm

\begin{equation}
\mathrm{a(x)}= 1 + 0.17699y \mathrm{-} 0.50447y^{2} \mathrm{-} 0.02427y^{3} + 0.72085y^{4} + 0.01979y^{5} \mathrm{-} 0.77530y^{6} + 0.32999y^{7},
\label{eq: eq. 2.13}
\end{equation}
\vskip 0.25cm

\begin{equation}
\rm{b(x)}= 1.41338y + 2.28305y^{2} + 1.07233y^{3} \mathrm{-} 5.38434y^{4}
\mathrm{-} \,0.62251y^{5} + 5.30260y^{6} \mathrm{-} \,2.09002y^{7},
\label{eq: eq. 2.14}
\end{equation}
\vskip 0.25cm

\ \\
\noindent
after obtaining the total theoretical spectral flux from 1090 wavelength data points ranging from 1005 \AA{} to 1600000 \AA{}, an interpolation was performed using equations \ref{eq: eq. 2.15} and \ref{eq: eq. 2.16} to find the exact theoretical photometric points and compare them with those obtained from the SED of VizieR\footnote{\url{http://vizier.cds.unistra.fr/vizier/sed/}}, which are listed in Table \ref{Tab: Tab. 2.7}, resulting in 1598996 interpolated data points in steps of 1 \AA{}ngstr\"om.

\begin{equation}
\mathrm{\lambda_{\textrm{intermediate}, j} = \lambda_{1} + j \cdot 1 \,\textrm{\AA}},
\label{eq: eq. 2.15}
\end{equation}
\vskip 0.25cm

\begin{equation}
\mathrm{F_{\textrm{interpolation}} (\lambda_{\textrm{intermediate}, j}) = F_{i} + \frac{(F_{i+1} - F_{i})}{(\lambda_{i+1} - \lambda_{i})} \cdot (\lambda_{\textrm{intermediate}, j} - \lambda_{i})},
\label{eq: eq. 2.16}
\end{equation}
\vskip 0.25cm

\ \\
\noindent
With the observed flux data from VizieR\footnote{\url{http://vizier.cds.unistra.fr/vizier/sed/}} and theoretical flux from VOSA\footnote{\url{http://svo2.cab.inta-csic.es/theory/newov2/index.php?models=Kurucz2003all}}, using the statistical chi-square $\mathrm{X^2}$ method with 7 degrees of freedom, I proceeded to identify the lowest chi-square in the regions near the ultraviolet, which was $\mathrm{\chi^2=4.744}$, obtaining the physical parameters for the star HD\,72968 presented in Table \ref{Tab: Tab. 2.8}.

\vskip 0.25cm
\begin{table}[]
\caption{Parameters obtained through the chi-square mathematical method using theoretical spectra. Where $\alpha$ corresponds to the relative overabundance of alpha nuclear elements in a star compared to the observed proportion in the Sun and (*) represents fixed or default values.}
\centering
\begin{tabular}{l l}
\hline
\textrm{Parameters} & \textrm{Values} \\
\hline
\hline
$\rm{T_{eff}}$          & $\rm{10500 \pm 250\,K} $           \\
$\rm{log{g}}$           & $\rm{4.5  \pm 0.5\,dex} $           \\
$\rm{R_{1}}$        & $\rm{2.5  \pm 0.1} \, \rm{R_{\odot}} $         \\
$\rm{*d}$        & $\rm{ 107.531 \pm 1.135 \,{pc}}$        \\
$\rm{*E(B-V)}$         & $\rm {0.038 \pm 0.0006}$ \\
$*\alpha$                & $\rm{0.0}$                       \\
$\rm{*[M/H]}$            & $\rm{0.00}$                      \\
$\rm{\chi^2}$           & $\rm{4.744}$                    \\
\hline
\end{tabular}
\label{Tab: Tab. 2.8}
\end{table}

\begin{figure}[]
\centering
\includegraphics[width=0.8\textwidth,angle=0]{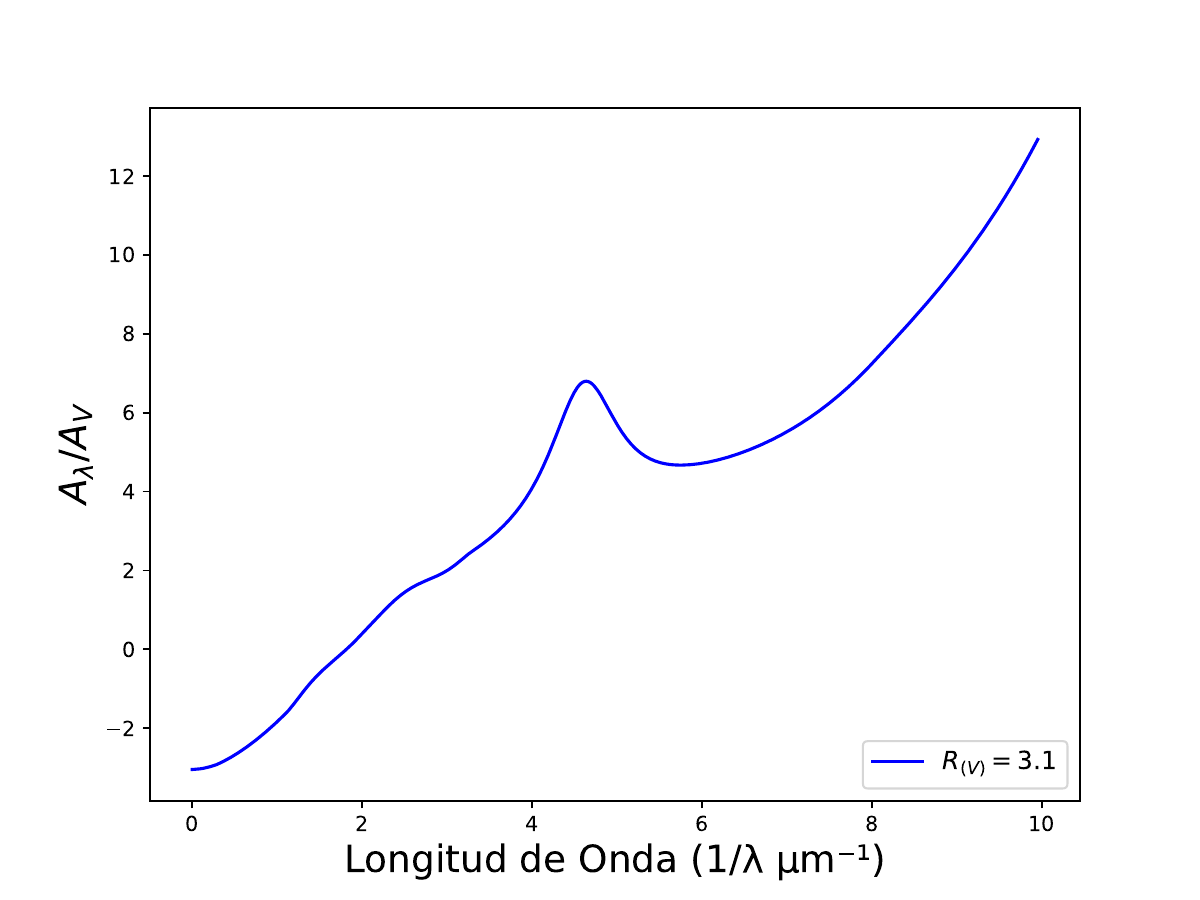}
\vskip 0.25cm
\caption{Extinction curve of $\mathrm{k(\lambda - V)}$ for wavelengths from $\mathrm{0.1 \,\mu m^{-1}}$ to $\mathrm{160\,\mu m^{-1}}$.}
\label{fig:Fig. 2.12}
\end{figure}

In Figure \ref{fig:Fig. 2.12}, the extinction curve of the values in equation \ref{eq: eq. 2.17} is expressed, following the considerations of \citet{1990ApJ...357..113M}, expressed in units of $\mathrm{1/\lambda \,\mu m^{-1} \,and\, A_{\lambda}/A_{V}}$, which corresponds to the specific extinction relative to the total extinction.

\begin{equation}
\mathrm{k(\lambda - V) = E(\lambda - V)/E(B - V)},
\label{eq: eq. 2.17}
\end{equation}

\begin{figure*}[]
\centering
\includegraphics[width=0.8\textwidth,angle=0]{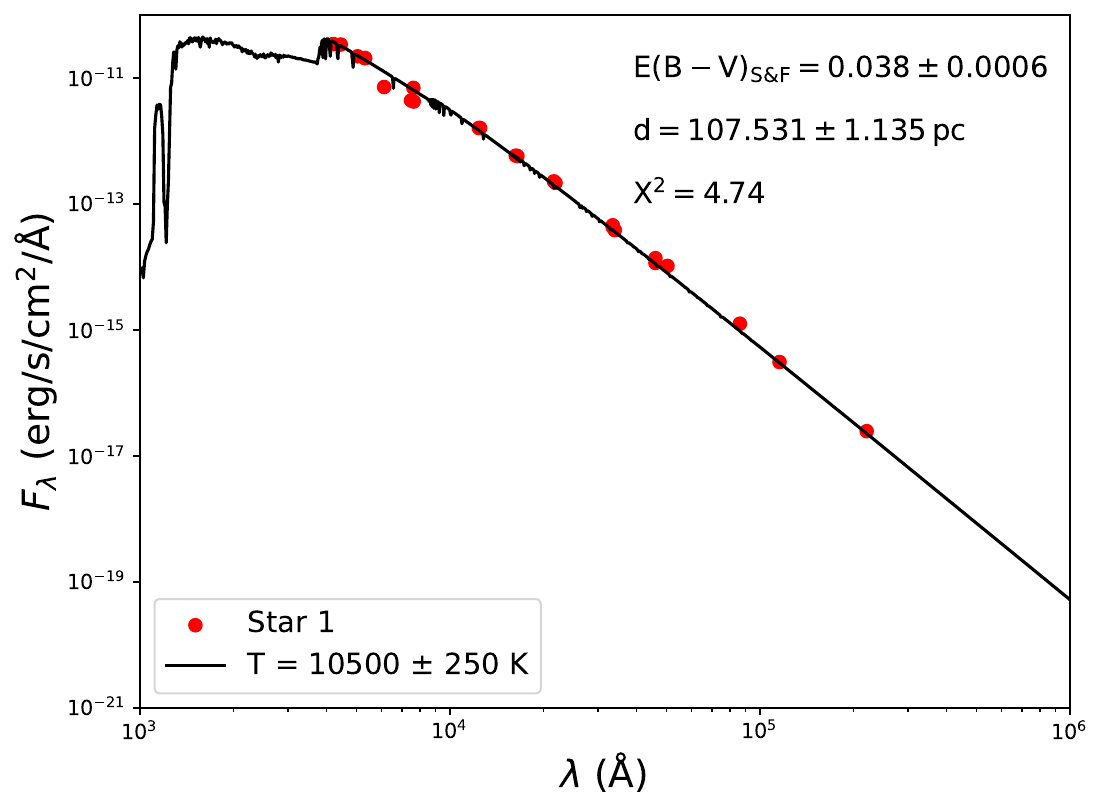}
\vskip 0.25cm
\caption{Theoretical SED spectral energy distribution for the star HD\,72968, with the physical parameters of $\mathrm{T_{eff} = \rm{10500 \pm 250\,K}}$, $\mathrm{log g = \rm{4.5 \pm 0.5\,dex}}$, $\mathrm{R_1=2.5 \pm 0.1 \,R_{\odot}}$ y $\mathrm{d=107.531 \pm 1.135 \,pc}$}
\label{fig:Fig. 2.13}
\end{figure*}

\newpage
\section{Conclusions}\label{Sec:Sec. 5}

In this research work, a photometric and spectroscopic analysis was carried out for the chemically peculiar star HD\,72968, where the following conclusions were made:

\begin{itemize}
\item For the photometric analysis, TESS (Transiting Exoplanet Survey Satellite) data were extracted, where the apparent magnitude of V = 5.6310 $\pm$ 0.0002 (mag) and the minimum magnitude of $m_{\mathrm{min}}$ = 5.6516 $\pm$ 0.0002 (mag) were obtained through the graphical representation of the magnitude histogram for HD\,72968. Using the TESS data, the PDM-IRAF algorithm was implemented to determine the period of the star HD\,72968, which resulted in P = 11.307 $\pm$ 0.005\,d, and the value of the apparent minimum magnitude at the time BJD 2976.14736220 was confirmed.
\item For a better graphical representation of the TESS photometric data, the wavelet data analysis technique was used, where we report a sub-period of 2.7 days possibly related to the star's pulsations, which is approximately ~76\% shorter than the observed period of 11.307 d.
\item We also identified changes in the morphology of the light curve at BJD=2976.15 and BJD=2963.31, where a sudden drop in the star's brightness was observed, possibly related to some ejection of material caused by pulsation.
\item From BJD=2963.31, it was observed that a disturbance in the light curve during the minimum is maintained, which is more extended and seems to be a remnant of the second eclipse or material ejection from the star.
\item Additionally, an estimation of the effective temperature was made through the colorimetric method, which included a color\textrm{-}color plot of the J\textrm{-}H and H\textrm{-}K bands for 27 spectral type variants from O9 = 35900 K to K4 = 4560 K. The position of the star HD\,72968 was identified with respect to other spectral types, obtaining the approximate effective temperature of 13550 $\pm$ 550\,K, a value that shows significant variations compared to the real effective temperature values due to the comparison with main-sequence stars and the anomaly in its chemical composition for HD\,72968.
\item For the spectroscopic analysis, UVES spectra were obtained from the ESO (European Southern Observatory) database. Spectral treatments such as trimming, normalization, and Doppler effect correction were performed using the IRAF software. The identification of the main chemical elements present in HD\,72968, such as Sr II, H $\mathrm{\delta}$, Cr II, and H $\mathrm{\beta}$, was carried out, each with its corresponding wavelength, spectral type, and equivalent width.
\item For the chemical abundance analysis of HD\,72968, the chemical elements and their atomic numbers were obtained from VizieR using the wavelengths of the main absorption lines. Subsequently, key parameters were determined using the NIST Atomic Spectra Database Lines Data to construct Table \ref{Tab: Tab. 2.10}. Finally, the SPECTRUM ABUNDANCE software was used to identify the most abundant chemical elements in the star HD\,72968, as presented in Figure \ref{fig:Fig. 2.14}.
\item From the spectral analysis, a theoretical spectrum was generated using the SPECTRUM software, considering the ATLAS9 stellar atmosphere models over a wide range of temperatures from 7000 K to 15000 K. Using the chi-square statistical method, the best theoretical spectrum for HD\,72968 was determined, obtaining the following parameters: $\mathrm{T_{eff} = 10250 \pm 250 \,K}$, where the theoretical effective temperature value closely matches those determined by other authors over time, $\mathrm{logg = 4.0 \pm 0.5 \,dex}$, $\mathrm{Vmicro = 0.0 \,kms^{-1}}$, $\mathrm{v_{sin i}= 8 \pm 1 \,km s^{-1}}$, $\mathrm{l/H}$ = 1.25, $\mathrm{M/H = 0.00}$, and $\mathrm{\chi^2 = 18.5}$. It should be noted that the chi-square value was calculated entirely between the observed and theoretical spectrum for wavelengths from 4800 \AA{} to 4900 \AA{}.
\item Additionally, a comparison was made between the star HD\,72968 and chemically peculiar stars of similar spectral type, namely HD\,220825 with $\mathrm{T_{eff} = 9250 K}$ and HD\,223640 with $\mathrm{T_{eff} = 12340 K}$. The spectra were extracted from the ESO database, and the respective treatments of trimming, normalization with the IRAF Chebyshev function, and Doppler effect correction were applied. The spectral comparison between HD\,72968 and HD\,220825 yielded a chi-square value of $\mathrm{\chi^2=34.3}$, and the spectral comparison between HD\,72968 and HD\,223640 yielded a chi-square value of $\mathrm{\chi^2=66.1}$. Both values were used as upper and lower limits for estimating the effective temperature.
\item A first spectral energy distribution (SED) was performed with the Kurucz ODFNEW / NOVER (2003) model in VOSA, obtaining the physical parameters for the star HD\,72968: $\mathrm{T_{eff} = 10000 K}$, $\mathrm{logg = 4.5 \pm \,dex}$, $\mathrm{metallicity = 0.5}$, $\mathrm{alpha = 0.0}$, and $\mathrm{Av = 0.195}$. Knowing the stellar parameters of HD\,72968 and observing that the temperature obtained in VOSA falls within the range of the effective temperature determined from the theoretical spectrum, the theoretical effective temperature was calculated using the data obtained from the SED in VOSA with equation \ref{eq: eq. 2.7}, yielding $\mathrm{T_{eff} = 10000 K}$.
\item To determine the stellar radius of HD\,72968, a theoretical spectral energy distribution (SED) was performed using the Kurucz ODFNEW / NOVER (2003) theoretical models from VOSA. Using equation \ref{eq: eq. 2.8}, the total theoretical spectral flux for HD\,72968 was determined for different wavelengths. Using the chi-square statistical method $\mathrm{\chi^2 = 4.744}$, the best theoretical model fitting the photometric data obtained from VIZIER-SED and spectroscopic analysis was determined, yielding the physical parameters, including the radius of HD\,72968 in solar radii: $\mathrm{T_{eff} = 10500 \pm 250\,K}$, $\mathrm{logg = 4.5 \pm 0.5\,dex}$, stellar radius $\mathrm{R_{1}=2.5 \pm 0.1 \, R_{\mathrm{\odot}}}$, and $\mathrm{d=107.531 \pm 1.135 \,pc}$.
\end{itemize}

\end{document}